\documentstyle[11pt,epic,eepic]{article}
\topmargin - 1.0 true cm
\oddsidemargin - 0.25 true cm
\newcommand{\beq}{\begin{equation}}
\newcommand{\eeq}{\end{equation}}
\def\bea{\begin{eqnarray}}
\def\eea{\end{eqnarray}}
\def\nn{\nonumber}
\def\sss{\scriptscriptstyle}

\def\barp{{\raise.35ex\hbox
{${\sss (}$}}---{\raise.35ex\hbox{${\sss )}$}}}
\def\bdbarp{\hbox{$B_d$\kern-1.4em\raise1.4ex\hbox{\barp}}}
\def\bsbarp{\hbox{$B_s$\kern-1.4em\raise1.4ex\hbox{\barp}}}

\def\roughly#1{\mathrel{\raise.3ex\hbox
{$#1$\kern-.75em\lower1ex\hbox{$\sim$}}}}

\def\Bstt{B \rightarrow X_s \tau^+ \tau^-} 
\def\Bsll{B \rightarrow X_s l^+ l^-} 
\def\Bsga{B \rightarrow X_s \gamma}
\begin{document}
\baselineskip=7truemm
\begin{flushright}
{\bf hep-ph/9908229} \\
HUPD-9910 \\
YUMS 99-020 \\
KEK-TH-633 \\
\end{flushright}

\begin{center}
\bigskip

{\Large \bf A Systematic Analysis of the Lepton Polarization
Asymmetries in the Rare $B$ Decay, $\Bstt$ }\\ 
\bigskip\bigskip

S. Fukae~$^{a,}$\footnote{fukae@ipc.hiroshima-u.ac.jp},~~
C. S. Kim~$^{b,}$\footnote{kim@cskim.yonsei.ac.kr,~~
http://phya.yonsei.ac.kr/\~{}cskim/ },~~ 
and~~ T. Yoshikawa~$^{c,}$\footnote{JSPS Research Fellow,~~
yosikawa@acorn.kek.jp}
\end{center}


\begin{flushleft}
~~~~~~~~~~~~~$a$: {\it Department of Physics, Hiroshima University,
 Higashi Hiroshima 739-8526, Japan}
\\
~~~~~~~~~~~~~$b$: {\it Department of Physics, Yonsei University, 
Seoul 120-749, Korea}\\
~~~~~~~~~~~~~$c$: {\it Theory Group, KEK, Tsukuba, Ibaraki 305-0801, Japan}
\end{flushleft}

\begin{center}

(\today )

\bigskip\bigskip

{\bf Abstract}

\end{center}

\begin{quote}
The most general model--independent analysis of the lepton polarization 
asymmetries in the rare $B$ decay, 
$\Bstt$, is presented. We present the longitudinal, normal and transverse 
polarization asymmetries for the $\tau^+ $ and $\tau^-$, and combinations
of them, as functions of the Wilson coefficients of  
twelve independent four--Fermi interactions, ten of them local and
two nonlocal.
These procedures will tell us which type of operators contributes to
the process. 
And it will be  very useful to pin down new physics systematically,
once we have the experimental data with high statistics
and a deviation from the Standard Model is found. 
\end{quote}
\newpage

\section{Introduction}

Rare $B$-meson decays are very useful for constraining new physics beyond 
the Standard Model (SM). 
In particular, the processes $\Bsga$ and $\Bsll$ are 
experimentally clean, and are possibly the most
sensitive to the various extensions of the SM because these decays occur
only through loops in the SM. 
Nonstandard model effects can manifest themselves in these rare decays 
through the Wilson coefficients, 
which can have values distinctly different from their
Standard Model counterparts. 
(See for example,
\cite{Goto,Wells,handoko,Wyler,Grossman,Rizzo,Jang,Cho,KKL})
Compared to $\Bsga$,  
the flavor changing leptonic decay $\Bsll$ is more sensitive to the actual
form of the new interactions since we can measure experimentally
various kinematical distributions as well as a total rate.
While new physics will change only the systematically uncertain normalization
for $\Bsga$, the interplay of various operators will also change the spectra
of the decay $\Bsll$. 

We can expect that the lepton polarization asymmetries in $\Bstt $ decay 
may also give useful information to fit parameters in the SM 
and constrain new physics \cite{Hewett,Kruger,HZ,GN,KKL,CGG}. 
We note that the previous studies for lepton flavor asymmetries
have been limited only to the subset of ten local four--Fermi
interactions within specific extended models,  such as
the two-Higgs-doublet model, the minimal supersymmetric model, 
the left-right symmetric model, {\it etc}. 
(see for example \cite{Wyler,AGM,Hewett}.)
In the SM and in many of its extensions, the decay $\Bsll$ is 
completely determined phenomenologically 
by the numerical values of Wilson coefficients
of only three operators evaluated at the scale $\mu \sim m_b$.
However, it would be most interesting if
this three--parameter fit were found  
unsuccessful to explain the real experimental distributions, 
and if the new interactions even necessarily implied
an extension 
of the ten local four--Fermi operator basis to 
new operators beyond the usual set \cite{Rizzo}. 
And, therefore, the new physics scenario can be much richer
than any of those models. 

In our previous work \cite{FKMY}, we studied the dependence 
on the four--Fermi interactions to the decay distribution and 
the forward-backward asymmetry of $B\rightarrow X_s l^+ l^-$, where $l$ 
is electron or muon \cite{ATM}. We also studied the correlation between the
branching ratio and the forward-backward asymmetry by changing each
coefficient. From the study we also found that the dependence 
of the correlation on the
coefficients of the vector--type interaction is
large and we can use such information to 
identify the corresponding new physics contribution.
However, we cannot get enough information for the scalar--
and tensor--type interactions from 
such correlation studies.
As the next step, we study here the case of $\Bstt $ and discuss the
importance of measuring the $\tau$ polarization asymmetries to
investigate the scalar-- and tensor--type interactions. The contributions
from the scalar and tensor interactions will appear in the difference 
between the asymmetries for $\tau^-$ and $\tau^+$. 

The paper is organized as follows.
In Section 2, we show the most general four--Fermi
interactions and the decay distribution of $\Bstt $. 
In the previous work \cite{FKMY}, 
we analyzed the decay $B \to X_s l^+ l^-$
based on ten local operators expansion. 
Here we expand our investigation by including the contribution
from two nonlocal type operators. 
In Section 3, we present the
longitudinal, normal and transverse
polarization asymmetries of $\tau^+ $ and $\tau^- $, and
study their dependence on the four--Fermi interactions. 
We discuss the difference between
the asymmetries for $\tau^-$ and $\tau^+$ in Section 4.  
The correlations between the branching ratio
and the asymmetries are discussed in Section 5. 
Section 6
summarizes the results.

\section{Dilepton Invariant Mass Distribution }

We follow  Refs. \cite{Kruger,FKMY,AGHM,KMS,Long} for notations and
for the choice of the parameters in the SM as well as for the 
incorporation of the long-distance effects of charmonium states. 
We start by defining the various kinematic variables.
In this paper, the inclusive semileptonic
$B$ decay is modeled by the partonic calculation, {\it i.e.},
$b(p_b) \rightarrow s(p_s) + l^+(p_+) + l^-(p_-) $. 
This is regarded as the leading order calculation in the $1/m_b$
expansion \cite{AGHM,Falk}.  
Then the decay distribution is described  
by the following two kinematic variables $s$ and $u$, 
\bea
s &=& ( p_b - p_s )^2 = ( p_+ + p_- )^2  
         = m_b^2 + m_s^2 + m_+^2 + m_-^2 - t_+ - t_- ,\nn \\
u &=& t_+ - t_- , \\
{\rm with}&&t_+ = ( p_s + p_+ )^2 = ( p_b - p_- )^2, \nn \\
&&t_- = ( p_s + p_- )^2 = ( p_b - p_+ )^2. \nn 
\eea
In the center of mass frame of the dileptons,
$u$ is written in terms of 
$\theta $, {\it i.e.}, the angle between the momentum of the $B$ meson
and that of $l^+$,
\bea
u &=& - u(s) \cdot cos\theta \equiv - u(s) z, \\  
{\rm with}&&~~~~z = \cos\theta,\nn \\
&&u(s) = \sqrt{({s-(m_b+m_s)^2})({s-(m_b-m_s)^2})(1 -
                            \frac{4m_l^2}{s})}. \nn
\eea
The phase space is defined in terms of  $s$ and $z$, 
\bea
4 m_l^2 \leq &s&  \leq ( m_b - m_s)^2, \nn \\
-1 \leq &z& \leq 1.
\eea

We want to consider the inclusive lepton polarization asymmetries
as functions of the Wilson coefficients of  
the following twelve most general independent four--Fermi interactions.
There are two nonlocal ($C_{SL},C_{BR}$), and ten local 
four--Fermi interactions;
\bea
{\cal M} = \frac{G_F~ \alpha }{\sqrt{2}\pi }~V_{ts}^*V_{tb} 
              &[& C_{SL} ~\bar{s} i \sigma_{\mu \nu } \frac{q^\nu}{q^2} 
                        ( m_s L ) b 
                         ~\bar{l} \gamma^\mu l \nn \\
              &+& C_{BR} ~\bar{s} i \sigma_{\mu \nu } \frac{q^\nu}{q^2} 
                        ( m_b R ) b 
                        ~\bar{l} \gamma^\mu l  \nn \\
              &+& C_{LL} ~\bar{s}_L \gamma_\mu b_L 
                           ~\bar{l}_L \gamma^\mu l_L  \nn \\
              &+& C_{LR} ~\bar{s}_L \gamma_\mu b_L  
                           ~\bar{l}_R \gamma^\mu l_R  \nn \\
              &+& C_{RL} ~\bar{s}_R \gamma_\mu b_R 
                           ~\bar{l}_L \gamma^\mu l_L  \nn \\
              &+& C_{RR} ~\bar{s}_R \gamma_\mu b_R  
                           ~\bar{l}_R \gamma^\mu l_R  \nn \\
              &+& C_{LRLR} ~\bar{s}_L b_R ~\bar{l}_L l_R \nn \\
              &+& C_{RLLR} ~\bar{s}_R b_L ~\bar{l}_L l_R \nn \\
              &+& C_{LRRL} ~\bar{s}_L b_R ~\bar{l}_R l_L \nn \\  
              &+& C_{RLRL} ~\bar{s}_R b_L ~\bar{l}_R l_L \nn \\
              &+& C_T       ~\bar{s} \sigma_{\mu \nu } b 
                           ~\bar{l} \sigma^{\mu \nu } l \nn \\
              &+& i C_{TE}   ~\bar{s} \sigma_{\mu \nu } b 
                           ~\bar{l} \sigma_{\alpha \beta } l 
                           ~\epsilon^{\mu \nu \alpha \beta } ].
\eea
where $C_{XX}$'s are the coefficients of
the four--Fermi interactions. 
Among them, there are two nonlocal four-Fermi interactions denoted by
$C_{SL}$ and $C_{BR}$, which correspond to $-2 C_7 $ in the SM, and which are
constrained by the experimental data of $b\rightarrow s \gamma $. 
There are four vector--type interactions denoted by
$C_{LL}$, $C_{LR}$,   $C_{RL}$, and  $C_{RR}$.
Two of them ($C_{LL}$, $C_{LR}$) are already present in the
SM as the combinations of  ($C_9-C_{10}$, $C_9+C_{10}$).
Therefore, they are regarded as the sum of the contribution from the SM
and the new physics deviations $( C_{LL}^{\rm new}, C_{LR}^{\rm new} )$.
The other vector interactions denoted by  $C_{RL}$ and $C_{RR}$
are obtained by interchanging  the chirality projections
$L \leftrightarrow R $.
There are four scalar--type interactions, $C_{LRLR}$, $C_{RLLR}$,
$C_{RLLR}$ and  $C_{RLRL}$. The remaining two denoted by $C_T$
and  $C_{TE}$ correspond to tensor--type. The subindices, $L$ and $R$, 
are chiral projections, 
$L =\frac{1}{2}( 1 - \gamma_5 ) $ and $R =\frac{1}{2}( 1 + \gamma_5 )$.

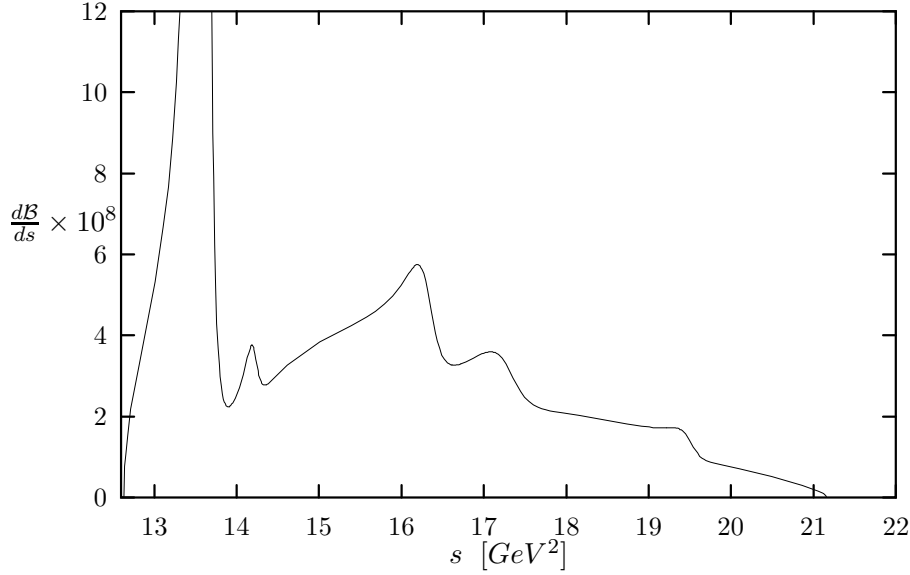
\begin{figure}[ht]
\setlength{\unitlength}{0.240900pt}
\begin{picture}(1500,900)(0,0)
\tenrm
\thicklines \path(220,113)(240,113)
\thicklines \path(1436,113)(1416,113)
\put(198,113){\makebox(0,0)[r]{0}}
\thicklines \path(220,240)(240,240)
\thicklines \path(1436,240)(1416,240)
\put(198,240){\makebox(0,0)[r]{2}}
\thicklines \path(220,368)(240,368)
\thicklines \path(1436,368)(1416,368)
\put(198,368){\makebox(0,0)[r]{4}}
\thicklines \path(220,495)(240,495)
\thicklines \path(1436,495)(1416,495)
\put(198,495){\makebox(0,0)[r]{6}}
\thicklines \path(220,622)(240,622)
\thicklines \path(1436,622)(1416,622)
\put(198,622){\makebox(0,0)[r]{8}}
\thicklines \path(220,750)(240,750)
\thicklines \path(1436,750)(1416,750)
\put(198,750){\makebox(0,0)[r]{10}}
\thicklines \path(220,877)(240,877)
\thicklines \path(1436,877)(1416,877)
\put(198,877){\makebox(0,0)[r]{12}}
\thicklines \path(272,113)(272,133)
\thicklines \path(272,877)(272,857)
\put(272,68){\makebox(0,0){13}}
\thicklines \path(401,113)(401,133)
\thicklines \path(401,877)(401,857)
\put(401,68){\makebox(0,0){14}}
\thicklines \path(530,113)(530,133)
\thicklines \path(530,877)(530,857)
\put(530,68){\makebox(0,0){15}}
\thicklines \path(660,113)(660,133)
\thicklines \path(660,877)(660,857)
\put(660,68){\makebox(0,0){16}}
\thicklines \path(789,113)(789,133)
\thicklines \path(789,877)(789,857)
\put(789,68){\makebox(0,0){17}}
\thicklines \path(919,113)(919,133)
\thicklines \path(919,877)(919,857)
\put(919,68){\makebox(0,0){18}}
\thicklines \path(1048,113)(1048,133)
\thicklines \path(1048,877)(1048,857)
\put(1048,68){\makebox(0,0){19}}
\thicklines \path(1177,113)(1177,133)
\thicklines \path(1177,877)(1177,857)
\put(1177,68){\makebox(0,0){20}}
\thicklines \path(1307,113)(1307,133)
\thicklines \path(1307,877)(1307,857)
\put(1307,68){\makebox(0,0){21}}
\thicklines \path(1436,113)(1436,133)
\thicklines \path(1436,877)(1436,857)
\put(1436,68){\makebox(0,0){22}}
\thicklines \path(220,113)(1436,113)(1436,877)(220,877)(220,113)
\put(42,545){\makebox(0,0)[l]{\shortstack{$ \frac{d {\cal B}}{d s} \times 10^8  $}}}
\put(828,23){\makebox(0,0){$ s ~~[GeV^2]$}}
\thinlines \path(224,113)(224,113)(225,162)(227,181)(230,211)(235,252)(248,322)(273,452)(286,538)(294,601)(301,677)(307,766)(310,832)(312,877)
\thinlines \path(362,877)(363,762)(364,686)(365,617)(367,511)(368,443)(370,386)(372,347)(374,321)(375,303)(377,288)(379,276)(380,268)(381,265)(382,263)(383,261)(384,259)(384,258)(385,257)(387,256)(388,256)(390,256)(391,257)(392,258)(393,260)(395,262)(399,269)(406,286)(412,306)(418,333)(420,340)(422,346)(423,349)(423,351)(425,353)(427,351)(428,349)(428,346)(430,339)(431,330)(435,312)(436,305)(438,299)(440,295)(440,293)(441,292)(442,291)(443,290)(445,290)(446,290)(448,290)
\thinlines \path(448,290)(449,291)(451,292)(457,298)(481,321)(532,358)(581,383)(605,397)(619,406)(633,417)(646,430)(659,446)(671,465)(674,469)(677,473)(679,475)(680,477)(681,478)(683,479)(684,479)(686,479)(688,478)(689,477)(690,475)(692,472)(695,465)(698,454)(701,438)(707,404)(713,372)(716,359)(720,347)(723,337)(725,334)(727,331)(730,327)(732,325)(733,324)(735,323)(737,322)(738,322)(739,322)(739,321)(740,321)(742,321)(743,321)(745,321)(746,322)(748,322)(750,322)(753,323)
\thinlines \path(753,323)(761,326)(773,332)(780,336)(786,339)(789,340)(792,341)(794,341)(795,341)(797,342)(798,342)(800,342)(802,342)(803,342)(804,341)(805,341)(807,341)(809,340)(812,338)(815,335)(818,332)(824,323)(835,301)(847,280)(853,271)(860,264)(867,259)(874,255)(880,253)(886,251)(893,249)(899,248)(913,246)(939,242)(989,233)(1014,229)(1037,225)(1049,224)(1055,223)(1058,223)(1062,223)(1063,223)(1065,223)(1066,223)(1067,223)(1068,223)(1070,223)(1071,223)(1073,223)(1074,223)(1076,223)
\thinlines \path(1076,223)(1078,223)(1080,223)(1081,223)(1082,223)(1082,223)(1083,223)(1085,223)(1087,223)(1088,223)(1090,223)(1091,222)(1093,222)(1094,222)(1096,221)(1097,220)(1099,220)(1101,218)(1103,216)(1106,213)(1112,203)(1118,192)(1125,183)(1128,178)(1132,175)(1135,173)(1138,171)(1145,169)(1151,167)(1164,164)(1191,158)(1241,146)(1289,132)(1314,123)(1321,120)(1323,119)(1324,118)(1325,117)(1326,116)(1327,114)
\end{picture}
\caption{The differential decay rate in the SM. }
\label{BRofBXstt}
\end{figure}

The differential decay rate of $B \rightarrow X_s l^+ l^- $ as a
function of the dilepton invariant mass is shown
as follows: 
\bea
\frac{d {\cal B}}{d s } = \frac{1}{2{m_b}^8}~{\cal B}_0 &{\rm Re}[& 
             S_1(s) ~\{ m_s^2 |C_{SL}|^2  + m_b^2 |C_{BR}|^2 \} \nn \\
           &+& S_2(s) ~\{ 2 m_b m_s C_{SL} C_{BR}^*\} \nn \\
           &+& S_3(s)~\{ 2 m_s^2 C_{SL} ( C_{LL}^* + C_{LR}^* )
                   + 2 m_b m_s C_{BR} ( C_{RL}^* + C_{RR}^*)  \}
                     \nn \\
           &+& S_4(s)~\{ 2 m_b^2 C_{BR} ( C_{LL}^* + C_{LR}^* )
                   + 2 m_b m_s C_{SL} ( C_{RL}^* + C_{RR}^* )  \}
                     \nn \\ 
           &+& S_5(s)~\{ 2 ( m_s C_{SL} 
                   +             m_b C_{BR} )C_{T}^* \} \nn \\
           &+& S_6(s)~\{ 4 (m_b C_{BR} - m_s C_{SL})C_{TE}^*
               \} \nn \\ 
             &+& M_2(s)~\{ { \left|C_{LL}\right|^2 + \left|C_{LR}\right|^2 
             + \left|C_{RL}\right|^2 + \left|C_{RR}\right|^2  } \} \nn \\
             &+& M_6(s)~\{ - 2 { ( C_{LL} C_{RL}^* 
                                      + C_{LR} C_{RR}^* )} \nn \\
             & & ~~~~~~ + {(C_{LRLR} C_{RLLR}^* 
                            + C_{LRRL} C_{RLRL}^* ) } \} \nn \\
             &+& M_8(s)~\{ { \left|C_{LRLR}\right|^2 
                                + \left|C_{RLLR}\right|^2 
                                + \left|C_{LRRL}\right|^2        
                                + \left|C_{RLRL}\right|^2  } \} \nn \\
             &+& M_9(s)~\{ 16 { \left|C_{T}\right|^2 +
                           64 
                                \left|C_{TE}\right|^2 
                                } \} \nn \\
             &+& N_1(s)~\{ 2 { ( C_{LL} C_{LR}^* 
                                      + C_{RL} C_{RR}^* )} \nn \\
             & & ~~~~~~ - {(C_{LRLR} C_{LRRL}^* 
                            + C_{RLLR} C_{RLRL}^* ) } \} \nn \\
             &+& N_5(s)~\{ - 2 {( C_{LL} C_{RL}^* 
                                      + C_{LR} C_{RR}^* )} \nn \\
             & & ~~~~~~ + {(C_{LRLR} C_{RLLR}^* 
                            + C_{LRRL} C_{RLRL}^* ) } \} \nn \\
             &+& N_5(s)~\{ 2 {( C_{LL} C_{RR}^* 
                                      + C_{LR} C_{RL}^* )} \nn \\
             & & ~~~~~~ + \frac{1}{2}{(C_{LRLR} C_{RLRL}^* 
                            + C_{LRRL} C_{RLLR}^* ) } \} \nn \\
             &+& N_6(s)~\{ 2 {( C_{LL} - C_{LR})(
                                      C_{LRLR}^* - C_{LRRL}^* )} \nn \\
             & & ~~~~~~ + 2 {(C_{RL} - C_{RR})(
                                 C_{RLLR}^* - C_{RLRL}^* ) } \} \nn \\
             &+& N_7(s)~\{ 2 {( C_{LL} - C_{LR})(
                                      C_{RLRL}^* - C_{RLLR}^* )} \nn \\
             & & ~~~~~~ + 2 {(C_{RL} - C_{RR})(
                                 C_{LRRL}^* - C_{LRLR}^* ) } \} \nn \\
             &+& N_8(s)~\{ 2 {( C_{LL} + C_{LR})(
                                      C_{T}^* )} 
                           + 2 {(C_{RL} + C_{RR})(
                                 C_{T}^* ) } \} \nn \\
              &+& N_9(s)~\{ 2 {( C_{LL} + C_{LR})(
                                      C_{TE}^* )} 
                           + 2 {(C_{RL} + C_{RR})(
                                 C_{TE}^* ) } \} \nn \\
             &+& N_5(s)~\{ {- 192 
                               \left|C_{TE}\right|^2 } \} ], 
\eea
where
${\cal B}_0$ is a normalization factor normalized to the semileptonic decay
\bea
{\cal B}_0 = {\cal B}_{sl}~ \frac{3 \alpha^2 }{16 \pi^2 }~
             \frac{ |V_{ts}^*V_{tb}|^2 }{|V_{cb}|^2 }~ 
            \frac{1}{f(\hat{m_c}) \kappa(\hat{m_c})} ,
\eea
and the phase space factor, $f(\hat{m_c}={m_c \over m_b})$, 
and the $O(\alpha_s)$
QCD correction factor \cite{Kim}, $\kappa(\hat{m_c})$, 
of $b \rightarrow c l \nu$ are given by
\bea
f(\hat{m_c}) &=& 1 - 8 \hat{m_c}^2 + 8 
        \hat{m_c}^6 - \hat{m_c}^8 - 24 \hat{m_c}^4 \ln \hat{m_c} ,  \\
\kappa(\hat{m_c}) &=& 1 - \frac{2 \alpha_s(m_b)}{3 \pi}~ 
           [(\pi^2-\frac{31}{4})(1-\hat{m_c})^2 + \frac{3}{2} ] . 
\eea
For the numerical calculations, we set  
$|V_{ts}^*V_{tb}|^2 / |V_{cb}|^2  =1$
and use the experimental value of the semileptonic branching fraction 
${\cal B}_{sl}=10.4 \%$. $S_n(s), M_n(s)$ and $N_n(s)$ are functions 
of the dilepton invariant mass $s$, 
\bea
S_1(s) &=& \frac{1}{s^2} u(s) [ 
               - 16 m_l^2 \{ (m_b^2 + m_s^2 ) s - ( m_b^2 - m_s^2 )^2\} 
               - 4 s \{ s^2 - \frac{1}{3} u(s)^2   
                       - ( m_b^2 - m_s^2 )^2  \}], \nn \\
S_2(s) &=& \frac{1}{s} u(s) m_b m_s ( - 32 m_l^2 - 16 s ), \nn \\
S_3(s) &=& \frac{1}{s} u(s) \{ 
                      8 m_l^2 ( s + m_b^2 - m_s^2 ) 
                     + 4 s ( s + m_b^2 -  m_s^2) \}, \nn \\
S_4(s) &=& \frac{1}{s} u(s) \{ 
                      8 m_l^2 ( s - m_b^2 + m_s^2 ) 
                     + 4 s ( s - m_b^2 +  m_s^2) \}, \nn \\
S_5(s) &=& \frac{1}{s} u(s) m_l\{
              96 m_bm_s s + 16 s ( m_b^2 + m_s^2 -s ) 
            + 32 ( s^2 - (m_b^2 - m_s^2 )^2 )\}, \nn \\
S_6(s) &=& \frac{1}{s} u(s) m_l\{
              96 m_bm_s s - 16 s ( m_b^2 + m_s^2 -s ) 
            - 32 ( s^2 - (m_b^2 - m_s^2 )^2 )\}, \nn \\
M_2(s) &=& 2 u(s) ( - \frac{1}{3} u(s)^2  - s^2 + ( m_b^2 - m_s^2)^2 ), \nn \\
M_6(s) &=& 8 u(s) m_b m_s (2m_l^2 + s ) ,  \nn \\
M_8(s) &=& - 2 u(s) ( m_b^2 + m_s^2 -s )( 2 m_l^2 - s ), \nn \\
M_9(s) &=& 2 u(s) \{ 4 m_l^2( m_b^2 - 6 m_b m_s + m_s^2 - s ) \nn \\
  & &~~~~ - \frac{2}{3} u(s)^2  - 2 ( m_b^2 + m_s^2) s 
      + 2  ( m_b^2 - m_s^2 )^2 \}, \nn \\
N_1(s) &=& 8 u(s) m_l^2 ( m_b^2 + m_s^2 -s ), \nn \\
N_5(s) &=& - 32 u(s) m_b m_s m_l^2, \nn \\
N_6(s) &=& 2 u(s) m_l m_b ( s - m_b^2 + m_s^2 ), \nn \\
N_7(s)&=& 2 u(s) m_l m_s ( s + m_b^2 - m_s^2 ), \nn \\
N_8(s) &=& 24 u(s) m_l ( - ( m_b + m_s ) s + (
                           m_b -  m_s ) (m_b^2 -  m_s^2 )), \nn \\
N_9(s) &=& 48 u(s) m_l ( ( m_b - m_s ) s - (
                           m_b + m_s ) (m_b^2 -  m_s^2 )). 
\eea
The terms with the functions $S_5(s),~S_6(s)$ and $N_n(s)$ appear only
in the case of massive leptonic decay, $\Bstt$.  For the massless cases,
$B\rightarrow X_s e^+ e^- $ and $B\rightarrow X_s \mu^+ \mu^- $, 
we already discussed 
the influence of new interactions on
the differential decay rate in Ref. \cite{FKMY}. 
Now we want to expand our discussion for the massive case, $\Bstt$. 
However, in this case it is not easy to 
determine 
the dependence of each coefficient of the twelve new interactions
from the given differential decay rate only, 
because of some terms with $N_n(s)$ which show the cross term dependence
between the different type operators, as shown in Eq. (5).
In the next Section we investigate instead the lepton polarization 
asymmetries. 
%
%
The dilepton's invariant mass distribution of $\Bstt$ within the SM is shown 
in Fig. \ref{BRofBXstt}

\section{Lepton Polarization Asymmetries}

We now compute the lepton polarization asymmetries from the four--Fermi
interactions defined in Eq. (4).  
We define the following orthogonal unit vectors, $S$ in the rest frame 
of $l^-$ and $W$ in the rest frame of $l^+$, 
for the polarization of the
leptons \cite{Hewett,Kruger} to the longitudinal direction ($L$),
the normal direction ($N$) and the transverse direction ($T$) 
\bea
S_{L}^{\mu} &\equiv& ( 0, {\bf e}_L ) = ( 0, \frac{{\bf p}_-}{|{\bf p}_-|} ),
                                           \nn \\
S_{N}^{\mu} &\equiv& ( 0, {\bf e}_N ) 
            = ( 0, \frac{{\bf p}_s \times {\bf p}_-}
                       {|{\bf p}_s \times{\bf p}_-|} ),
                                           \nn \\
S_{T}^{\mu} &\equiv& ( 0, {\bf e}_T ) = ( 0, {\bf e }_N\times {\bf e}_L ),
                                           \nn \\
W_{L}^{\mu} &\equiv& ( 0, {\bf w}_L ) = ( 0, \frac{{\bf p}_+}{|{\bf p}_+|} ),
                                           \nn \\
W_{N}^{\mu} &\equiv& ( 0, {\bf w}_N ) 
            = ( 0, \frac{{\bf p}_s \times {\bf p}_+}
                       {|{\bf p}_s \times{\bf p}_+|} ),
                                           \nn \\
W_{T}^{\mu} &\equiv& ( 0, {\bf w}_T ) = ( 0, {\bf w }_N\times {\bf w}_L ),
                                           \nn 
\eea
where ${\bf p}_\pm $ and ${\bf p}_s$ are the three momenta of the $l^\pm $ 
and the final strange ($s$) quark in the center--of--mass (CM) 
frame of the $l^+ l^- $ system. 
The longitudinal unit vectors, $S_L$ and $W_L $, are boosted 
by Lorentz transformation to CM frame of $l^+ l^-$, 
\bea
S_{L_{\rm CM}}^\mu &=& \left( \frac{|{\bf p}_-|}{m_l}, 
                    \frac{E_l {\bf p}_-}{ m_l |{\bf p}_-|
                     }\right), \nn \\
W_{L_{\rm CM}}^\mu &=& \left( \frac{|{\bf p}_-|}{m_l}, 
                    - \frac{E_l {\bf p}_-}{ m_l |{\bf p}_-|
                     }\right), \nn 
\eea 
while the vectors of perpendicular direction are not changed by the boost. 

The differential decay rate of $ B \rightarrow X_s l^+ l^- $ for any 
spin direction of leptons can be computed from the following 
spinor expression of the matrix elements:
\bea
{\cal M } &=& C_{SL} ~\bar{u}(p_s) i \sigma_{\mu \nu } \frac{q^\nu}{q^2} 
                        ( m_s L ) u(p_b) 
                         ~\bar{u}(p_-) P \gamma^\mu Q v(p_+)  \nn \\
              &+& C_{BR} ~\bar{u}(p_s) i \sigma_{\mu \nu } \frac{q^\nu}{q^2} 
                        ( m_b R ) u(p_b) 
                        ~\bar{u}(p_-) P \gamma^\mu Q v(p_+)  \nn \\
              &+& C_{LL} ~\bar{u}(p_s) \gamma_\mu L u(p_b) 
                           ~\bar{u}(p_-) P \gamma^\mu L Q v(p_+)  \nn \\
              &+& C_{LR} ~\bar{u}(p_s) \gamma_\mu L u(p_b) 
                           ~\bar{u}(p_-) P \gamma^\mu R Q v(p_+)  \nn \\
              &+& C_{RL} ~\bar{u}(p_s) \gamma_\mu R u(p_b) 
                           ~\bar{u}(p_-) P \gamma^\mu L Q v(p_+)  \nn \\
              &+& C_{RR} ~\bar{u}(p_s) \gamma_\mu R u(p_b) 
                           ~\bar{u}(p_-) P \gamma^\mu R Q v(p_+)  \nn \\
              &+& C_{LRLR} ~\bar{u}(p_s) R u(p_b) 
                           ~\bar{u}(p_-) P R Q v(p_+)  \nn \\
              &+& C_{LRRL} ~\bar{u}(p_s) R u(p_b) 
                           ~\bar{u}(p_-) P L Q v(p_+)  \nn \\
              &+& C_{RLLR} ~\bar{u}(p_s) L u(p_b) 
                           ~\bar{u}(p_-) P R Q v(p_+)  \nn \\
              &+& C_{RLRL} ~\bar{u}(p_s) L u(p_b) 
                           ~\bar{u}(p_-) P L Q v(p_+)  \nn \\
              &+& C_T       ~\bar{u}(p_s) \sigma_{\mu \nu } u(p_b)
                           ~\bar{u}(p_-) P \sigma^{\mu \nu } Q v(p_+) \nn \\
              &+& i C_{TE}   ~\bar{u}(p_s) \sigma_{\mu \nu } u(p_b) 
                           ~\bar{u}(p_-) P \sigma_{\alpha \beta } Q v(p_+) 
                           ~\epsilon^{\mu \nu \alpha \beta },
\label{matrixelement}
\eea
where $P =\frac{1}{2} ( 1 + \gamma_5 \gamma_\mu S^\mu )$ and 
$Q =\frac{1}{2} ( 1 + \gamma_5 \gamma_\mu W^\mu )$ are 
spin projection operators. 
Then the lepton polarization asymmetries are defined as 
\bea
P_x^- &\equiv & \frac{( \frac{d\Gamma( S_x, W_x )}{ds} + 
                        \frac{d\Gamma( S_x, - W_x )}{ds} )
                       - ( \frac{d\Gamma( - S_x, W_x )}{ds}
                         + \frac{d\Gamma( - S_x, - W_x )}{ds} )}
                      {( \frac{d\Gamma( S_x, W_x )}{ds} + 
                        \frac{d\Gamma( S_x, - W_x )}{ds} )
                       + ( \frac{d\Gamma( - S_x, W_x )}{ds}
                         + \frac{d\Gamma( - S_x, - W_x )}{ds} )}, \\
P_x^+ &\equiv & \frac{( \frac{d\Gamma( S_x, W_x )}{ds} + 
                        \frac{d\Gamma( - S_x, W_x )}{ds} )
                       - ( \frac{d\Gamma( S_x, - W_x )}{ds}
                         + \frac{d\Gamma( - S_x, - W_x )}{ds} )}
                      {( \frac{d\Gamma( S_x, W_x )}{ds} + 
                        \frac{d\Gamma( S_x, - W_x )}{ds} )
                       + ( \frac{d\Gamma( - S_x, W_x )}{ds}
                         + \frac{d\Gamma( - S_x, - W_x )}{ds} )},
\eea
where the subindex $x$ is $L, T$ or $N$. $P^\pm $ denotes
lepton polarization asymmetry of charged lepton $l^\pm$. 
$P_L$ denotes the
longitudinal polarization, $P_T$ is the polarization asymmetry in the
decay plane and $P_N$ is the normal component to both of them. 
As we can see from the direction of the lepton polarization, 
$P_L$ and $P_T$ are P-odd, T-even and CP-even observables 
while $P_N$ is P-even, T-odd and CP-odd observable\footnote{This is
because the time reversal operation changes 
the signs of momentum and spin, and because the parity transformation changes
only the sign of momentum.}.

The longitudinal polarization asymmetries for each lepton are
\bea
P_L^- = \frac{{\cal B}_0~ u(s) }{{m_b}^8 ~\frac{d {\cal B}}{d s}}&{\rm Re}[&
                 2 L_1(s)~\{  m_b^2 C_{BR} ( C_{LL}^*  
                             - C_{LR}^* )
                             + m_b m_s C_{SL} ( C_{RL}^* - C_{RR}^* ) 
                                                     \}    \nn \\
             &+& 2 L_2(s)~\{ - m_s^2 C_{SL} ( C_{LL}^* -
                                                C_{LR}^*  ) 
                              - m_b m_s C_{BR} ( C_{RL}^* - C_{RR}^* ) 
                                          \}        \nn \\ 
             &+& L_3(s)~\{ |C_{LL}|^2 - |C_{LR}|^2 + |C_{RL}|^2 - |C_{RR}|^2 
                            - 128 C_{T} C_{TE}^* 
                                                \} \nn \\
             &+& L_4(s)~\{ 2 C_{LL} C_{RL}^* - 2 C_{LR} C_{RR}^* 
                          - C_{LRLR} C_{RLLR}^* 
                          + C_{LRRL} C_{RLRL}^* \} \nn \\
             &+& 2 L_5(s)~\{ m_b C_{BR} ( - C_{T}^* + 2 C_{TE}^* ) 
                             + m_s C_{SL}( C_{T}^* + 2 C_{TE}^* )\} \nn \\ 
             &+& 2 L_6(s)~ m_b\{ (C_{LL} - C_{LR})
                                   ( C_{LRLR}^* + C_{LRRL}^* ) \nn \\ 
             & & ~~~~~~~~ + (C_{RL} - C_{RR})
                                   ( C_{RLLR}^* + C_{RLRL}^* ) \nn \\
             & & ~~~~~~~~ + 4 [ ( -3C_{LL} +C_{LR})
                                     ( C_T^*  - 2 C_{TE}^* ) \nn \\
             & & ~~~~~~~~ + ( - C_{RL} + 3 C_{RR} )
                                     ( C_T^*  + 2 C_{TE}^* )]
                                                            \} \nn \\
             &+& 2 L_7(s)~ m_s\{ (C_{LL} - C_{LR})
                                   ( C_{RLLR}^* + C_{RLRL}^* ) \nn \\ 
             & & ~~~~~~~~ + (C_{RL} - C_{RR})
                                   ( C_{LRLR}^* + C_{LRRL}^* ) \nn \\
             & & ~~~~~~~~ +
                      4 [(C_{LL}- 3 C_{LR} )( C_{T}^* + 2 C_{TE}^* )
                           \nn \\
             & & ~~~~~~~~ + (3C_{RL}-C_{RR})
                                    ( C_{T}^* - 2 C_{TE}^* )]
                                                            \} \\
             &+& L_8(s)~\{ - |C_{LRLR}|^2 + |C_{LRRL}|^2 
                            - |C_{RLLR}|^2 + |C_{RLRL}|^2 - 128 C_T C_{TE}^*
                                                            \}], \nn  
\eea
and
\bea
P_L^+ = \frac{{\cal B}_0~ u(s) }{{m_b}^8~ \frac{d {\cal B}}{d s}}&{\rm Re}[&
                 2 L_1(s)~\{  - m_b^2 C_{BR} ( C_{LL}^*  
                             - C_{LR}^* )
                             - m_b m_s C_{SL} ( C_{RL}^* - C_{RR}^* ) 
                                                     \}    \nn \\
             &+& 2 L_2(s)~\{ m_s^2 C_{SL} ( C_{LL}^* -
                                                C_{LR}^*  ) 
                              + m_b m_s C_{BR} ( C_{RL}^* - C_{RR}^* ) 
                                            \}      \nn \\ 
             &+& L_3(s)~\{ - |C_{LL}|^2 + |C_{LR}|^2 - |C_{RL}|^2 + |C_{RR}|^2 
                            - 128 C_{T} C_{TE}^* 
                                                \} \nn \\
             &+& L_4(s)~\{ - 2 C_{LL} C_{RL}^* + 2 C_{LR} C_{RR}^* 
                          - C_{LRLR} C_{RLLR}^* 
                          + C_{LRRL} C_{RLRL}^* \} \nn \\
             &+& 2 L_5(s)~\{ m_b C_{BR} ( - C_{T}^* + 2 C_{TE}^* ) 
                             + m_s C_{SL}( C_{T}^* + 2 C_{TE}^* )\} \nn \\ 
             &+& 2 L_6(s)~ m_b\{ (C_{LL} - C_{LR})
                                   ( C_{LRLR}^* + C_{LRRL}^* ) \nn \\
             & & ~~~~~~~~ + (C_{RL} - C_{RR})
                                   ( C_{RLLR}^* + C_{RLRL}^* ) \nn \\
             & & ~~~~~~~~ + 4 [ ( C_{LL} - 3 C_{LR})
                                     ( C_T^*  - 2 C_{TE}^* ) \nn \\
             & & ~~~~~~~~ +( 3C_{RL} - C_{RR} )
                                     ( C_T^*  + 2 C_{TE}^* )]   
                                                            \} \nn \\
             &+& 2 L_7(s)~ m_s\{ (C_{LL} - C_{LR})
                                   ( C_{RLLR}^* + C_{RLRL}^* ) \nn \\
             & & ~~~~~~~~ + (C_{RL} - C_{RR})
                                   ( C_{LRLR}^* + C_{LRRL}^* ) \nn \\
             & & ~~~~~~~~ + 4 [(- 3 C_{LL}+ C_{LR} )
                                          ( C_{T}^* + 2 C_{TE}^* ) \nn 
                                                            \\
              & & ~~~~~~~ + (- C_{RL} + 3 C_{RR})
                                    ( C_{T}^* - 2 C_{TE}^* )]
                                                            \} \\
             &+& L_8(s)~\{ - |C_{LRLR}|^2 + |C_{LRRL}|^2 
                            - |C_{RLLR}|^2 + |C_{RLRL}|^2 - 128 C_T C_{TE}^*
                                                            \}], \nn  
\eea
where $L_n(s)$ are the functions of kinetic variables $s$
\bea
L_1(s) &=& 2 ( - s + m_b^2 - m_s^2 ) v(s), \nn \\
L_2(s) &=& 2 ( s + m_b^2 - m_s^2 ) v(s), \nn \\
L_3(s) &=& \{ ( s^2 - ( m_b^2 -m_s^2 )^2 )v(s) 
                    + u(s)^2 \frac{1}{3 v(s)} \}, \nn \\
L_4(s) &=& 4 m_b m_s s v(s), \nn \\
L_5(s) &=& \frac{8}{s} m_l \{ - ( m_b^2 -m_s^2 )^2 v(s) 
                    + s ( m_b -m_s )^2 v(s)
                    + u(s)^2  \frac{1}{3 v(s)} \}, \nn \\
L_6(s) &=& m_l ( - s + m_b^2 - m_s^2 ) v(s), \nn \\
L_7(s) &=& m_l ( s + m_b^2 - m_s^2 ) v(s), \nn \\
L_8(s) &=& s ( m_b^2 + m_s^2 - s ) v(s),  \label{FuncL} \\
{\rm with}&&v(s) = \sqrt{ 1 - \frac{4 m_l^2 }{s}}. \nn 
\eea
We note that the contributions from the SM ({\it i.e.} from $C_{BR},
C_{SL}, C_{LL}$ and $C_{LR}$)
to $P_L^-$ are exactly the same as those to
$P_L^+$, but with the opposite sign. However, the contributions from new
interactions are totally different. This difference is very interesting and
useful for the search for the new physics effects later. 

The transverse asymmetries, $P_T^-$ and $P_T^+$, are; 
\bea
P_T^- = \frac{ {\cal B}_0~ u(s) w(s)}{{m_b}^8~ \frac{d {\cal B}}{d s}}
   &{\rm Re}[&  T_1(s)~\{  - m_b^2 |C_{BR}|^2 + m_s^2 |C_{SL}|^2 \}
                                                         \nn \\
             &+& 4 m_l~\{ m_b^2 C_{BR} ( 3 C_{LL}^* + C_{LR}^* ) 
                            + m_s^2 C_{SL} (C_{LL}^* + 3 C_{LR}^* ) 
                                                  \nn \\ 
             & & ~~~~ - m_b m_s C_{BR} ( 3 C_{RL}^* + C_{RR}^* )
                               -m_b m_s C_{SL} ( C_{RL}^* + 3 C_{RR}^* )
                            \} \nn \\
             &+& T_2(s)~\{ - |C_{LL}|^2 + |C_{RR}|^2 
                                                \} \nn \\
             &+& 2 T_3(s)~\{ - C_{LL} C_{LR}^* + C_{RL} C_{RR}^* \nn \\
             & & ~~~~ + ( 2 C_{T} - 4 C_{TE}) 
                                 ( C_{LRLR}^* - C_{LRRL}^* ) \nn \\
             & & ~~~~ +
                              ( 2 C_{T} + 4 C_{TE}) 
                                        ( C_{RLLR}^* - C_{RLRL}^* )
                                               \} \nn \\
             &+& T_4(s)~\{ |C_{LR}|^2 - |C_{RL}|^2 \} \nn \\
             &+& 2 T_5(s)~\{ m_b C_{BR} ( - C_{LRLR}^* + C_{LRRL}^* )
                            + m_s C_{SL} ( - C_{RLLR}^* + C_{RLRL}^* )
                                                            \} \nn \\
             &+& 2 T_6(s)~\{ m_b C_{BR}(C_{T}^* - 2 C_{TE}^*)
                             - m_s C_{SL} ( C_{T}^* + 2 C_{TE}^* )
                                                            \} \nn \\
             &+& 4 m_l^2~\{ m_b ( C_{LL} C_{LRLR}^* - C_{LR} C_{LRRL}^*)
                           +  m_s ( C_{LL} C_{RLLR}^* - C_{LR} C_{RLRL}^*) 
                                  \nn \\
             & & ~~~~ + m_s ( C_{RL} C_{LRLR}^* - C_{RR} C_{LRRL}^*)
                           +   m_b ( C_{RL} C_{RLLR}^* - C_{RR} C_{RLRL}^*)
                                                            \nn \\ 
             & & ~~~~ - 12 m_b ( C_{T} - 2 C_{TE}) 
                                            ( C_{LL}^* - C_{RR}^* ) \nn \\
            & & ~~~~ - 12 m_s (C_{T} + 2 C_{TE})
                                          ( C_{LR}^* - C_{RL}^* )
                                                      \} \nn \\
             &+&  2 T_7(s)\{ m_b ( C_{LL} C_{LRRL}^* - C_{LR}
                                                        C_{LRLR}^* )
                           +  m_s ( C_{LL} C_{RLRL}^* - C_{LR}
                                                        C_{RLLR}^* ) 
                                  \nn \\
             & & ~~~~ + m_s ( C_{RL} C_{LRRL}^* - C_{RR} C_{LRLR}^*)
                           +   m_b ( C_{RL} C_{RLRL}^* - C_{RR} C_{RLLR}^*)
                                                            \nn \\ 
             & & ~~~~ + 4 m_s ( C_{T} + 2 C_{TE} )
                                         ( C_{LL}^* - C_{RR}^* ) 
                           +   4 m_b ( C_{T} - 2 C_{TE} )
                                        ( C_{LR}^* - C_{RL}^* ) 
                          \} \nn \\
             &+& 256 m_l ( m_b^2 - m_s^2 ) C_{T} C_{TE}^* ],
\eea
and
\bea
P_T^+ = \frac{ {\cal B}_0~  u(s) w(s)}{{m_b}^8~ \frac{d {\cal B}}{d s}}
     &{\rm Re}[&  T_1(s)~\{  - m_b^2 |C_{BR}|^2 + m_s^2 |C_{SL}|^2 \}
                                                         \nn \\
             &+& 4 m_l~\{ m_b^2 C_{BR} ( C_{LL}^* + 3 C_{LR}^* ) 
                            + m_s^2 C_{SL} (3 C_{LL}^* +  C_{LR}^* ) 
                                                  \nn \\ 
             & & ~~~~ - m_b m_s C_{BR} ( C_{RL}^* + 3 C_{RR}^* )
                               -m_b m_s C_{SL} ( 3 C_{RL}^* + C_{RR}^* )
                            \} \nn \\
             &+& T_2(s)~\{ - |C_{LL}|^2 + |C_{RR}|^2 
                                                \} \nn \\
             &+& 2 T_3(s)~\{ - C_{LL} C_{LR}^* + C_{RL} C_{RR}^* \nn \\
             & & ~~~~ - ( 2 C_{T} - 4 C_{TE} ) 
                                     ( C_{LRLR}^* - C_{LRRL}^*) \nn \\
             & & ~~~~ - ( 2 C_{T} + 4 C_{TE} )
                                        ( C_{RLLR}^* - C_{RLRL}^* )
                                               \} \nn \\
             &+& T_4(s)~\{ |C_{LR}|^2 - |C_{RL}|^2 \} \nn \\
             &+& 2 T_5(s)~\{ - m_b C_{BR} ( - C_{LRLR}^* + C_{LRRL}^* )
                            - m_s C_{SL} ( - C_{RLLR}^* + C_{RLRL}^* )
                                                            \} \nn \\
             &+& 2 T_6(s)~\{ m_b C_{BR}(C_{T}^* - 2 C_{TE}^*)
                             - m_s C_{SL} ( C_{T}^* + 2 C_{TE}^* )
                                                            \} \nn \\
             &+& 4 m_l^2~\{ m_b ( C_{LL} C_{LRRL}^* - C_{LR} C_{LRLR}^*)
                           +  m_s ( C_{LL} C_{RLRL}^* - C_{LR} C_{RLLR}^*) 
                                  \nn \\
             & & ~~~~ + m_s ( C_{RL} C_{LRRL}^* - C_{RR} C_{LRLR}^*)
                           +   m_b ( C_{RL} C_{RLRL}^* - C_{RR} C_{RLLR}^*)
                                                            \nn \\ 
             & & ~~~~ - 12 m_s ( C_{T} + 2 C_{TE}) 
                                            ( C_{LL}^* - C_{RR}^* ) \nn \\
            & & ~~~~ - 12 m_b (C_{T} - 2 C_{TE})
                                          ( C_{LR}^* - C_{RL}^* )
                                                      \} \nn \\
             &+&  2 T_7(s)\{ m_b ( C_{LL} C_{LRLR}^* - C_{LR} C_{LRRL}^*)
                           +  m_s ( C_{LL} C_{RLLR}^* - C_{LR} C_{RLRL}^*) 
                                  \nn \\
             & & ~~~~ + m_s ( C_{RL} C_{LRLR}^* - C_{RR} C_{LRRL}^*)
                           +   m_b ( C_{RL} C_{RLLR}^* - C_{RR} C_{RLRL}^*)
                                                            \nn \\ 
             & & ~~~~ + 4 m_b ( C_{T} - 2 C_{TE} )
                                         ( C_{LL}^* - C_{RR}^* ) 
                           +   4 m_s ( C_{T} + 2 C_{TE} )
                                        ( C_{LR}^* - C_{RL}^* ) 
                          \} \nn \\
             &+& 256 m_l  ( m_b^2 - m_s^2 ) C_{T} C_{TE}^* ],
\eea
where $T_n(s)$, $w(s)$ are the functions of kinetic variables $s$
\bea
T_1(s) &=& \frac{8}{s} m_l ( m_b^2 - m_s^2 ), \nn \\
T_2(s) &=&  2 m_l ( s + m_b^2 - m_s^2 ), \nn \\
T_3(s) &=& 2 m_l s, \nn \\
T_4(s) &=& 2 m_l ( - s + m_b^2 - m_s^2 ), \nn \\
T_5(s) &=& s, \nn \\
T_6(s) &=& 4 ( m_b^2 - m_s^2 ) ( 1 + \frac{4 m_l^2 }{s} ), \nn \\
T_7(s) &=& ( 2 m_l^2 - s ),  \label{FuncT} \\
{\rm and}~~~~~w(s) &=& \frac{\pi u(s) }{4 \sqrt{s} v(s) }. \nn 
\eea

Finally the normal asymmetries, $P_N^-$ and $P_N^+$, are;
\bea
P_N^- = - \frac{ {\cal B}_0~ u(s) z(s)}{{m_b}^8~ \frac{d {\cal B}}{d s}}
     &{\rm Im}[&  \frac{8 m_l}{s}~\{  m_b^2 C_{BR} ( C_{LL}^* - C_{LR}^* ) 
                           + m_s^2 C_{SL} ( C_{LL}^* - C_{LR}^* ) 
                                \nn \\ 
             & & ~~~~ + m_b m_s ( C_{BR} + C_{SL} )
                                 ( - C_{RL}^* + C_{RR}^* ) \}
                                                         \nn \\
             &+& 8 m_l~\{ C_{LL} C_{LR}^* - C_{RL} C_{RR}^* 
                                                  \nn \\
             & & ~~~~ + 2 (C_{T} - 2 C_{TE} )
                                   ( C_{LRLR}^* + C_{LRRL}^* ) \nn \\
             & & ~~~~ + 2 (C_{T} + 2 C_{TE} )
                                   ( C_{RLLR}^* + C_{RLRL}^* )
                            \} \nn \\
             &+& 4~\{ - m_b C_{BR}(C_{LRLR}^* + C_{LRRL}^* )
                             - m_s C_{SL}(C_{RLLR}^* + C_{RLRL}^* )
                                           \nn \\
            & & ~~~~ + C_{LL} ( m_b C_{LRRL}^* + m_s C_{RLRL}^* )
                              + C_{LR} (m_b C_{LRLR}^* + m_s C_{RLLR}^* )
                                               \nn \\
            & & ~~~~+ C_{RL} ( m_s C_{LRRL}^* + m_b C_{RLRL}^* )
                              + C_{RR} (m_s C_{LRLR}^* + m_b C_{RLLR}^* )
                                   \nn \\
            & & ~~~~ + 4 ( C_{T} + 2 C_{TE} )
                                    ( - m_s C_{LL}^* + m_b C_{RL}^* ) 
                               \nn \\
            & & ~~~~ + 4 ( C_{T} - 2 C_{TE} )
                                    ( m_b C_{LR}^* - m_s C_{RR}^* ) 
                                          \}     \nn \\
             &+& \frac{16}{s} (m_b^2 - m_s^2)
                 ~\{ m_b C_{BR} ( C_{T}^* -2 C_{TE}^* )
                            + m_s C_{SL} ( C_{T}^* +2 C_{TE}^* ) \}
            ],
\eea
\bea
P_N^+ = - \frac{  {\cal B}_0~ u(s) z(s)}{{m_b}^8~ \frac{d {\cal B}}{d s}}
    &{\rm Im}[&  \frac{8 m_l}{s}~\{  - m_b^2 C_{BR} ( C_{LL}^* - C_{LR}^* ) 
                           - m_s^2 C_{SL} ( C_{LL}^* - C_{LR}^* ) 
                                \nn \\ 
             & & ~~~~ - m_b m_s ( C_{BR} + C_{SL} )
                                 ( - C_{RL}^* + C_{RR}^* ) \}
                                                         \nn \\
             &+& 8 m_l~\{ - C_{LL} C_{LR}^* + C_{RL} C_{RR}^* 
                                                  \nn \\
             & & ~~~~ - 2 (C_{T} - 2 C_{TE} )
                                   ( C_{LRLR}^* + C_{LRRL}^* ) \nn \\
             & & ~~~~ - 2 (C_{T} + 2 C_{TE} )
                                   ( C_{RLLR}^* + C_{RLRL}^* )
                            \} \nn \\
             &+& 4 ~\{ m_b C_{BR}(C_{LRLR}^* + C_{LRRL}^* )
                             + m_s C_{SL}(C_{RLLR}^* + C_{RLRL}^* )
                                           \nn \\
            & & ~~~~ - C_{LL} ( m_b C_{LRLR}^* + m_s C_{RLLR}^* )
                              - C_{LR} (m_b C_{LRRL}^* + m_s C_{RLRL}^* )
                                               \nn \\
            & & ~~~~ - C_{RL} ( m_s C_{LRLR}^* + m_b C_{RLLR}^* )
                              - C_{RR} (m_s C_{LRRL}^* + m_b C_{RLRL}^* )
                                   \nn \\
            & & ~~~~ + 4 ( C_{T} -  2 C_{TE} )
                                    ( m_b C_{LL}^* - m_s C_{RL}^* ) 
                               \nn \\
            & & ~~~~ + 4 ( C_{T} + 2 C_{TE} )
                                    ( - m_s C_{LR}^* + m_b C_{RR}^* ) 
                                          \}     \nn \\
             &+& \frac{16}{s} (m_b^2 - m_s^2)
                ~\{ m_b C_{BR} ( C_{T}^* -2 C_{TE}^* )
                            + m_s C_{SL} ( C_{T}^* +2 C_{TE}^* ) \}
            ],
\eea
where 
\bea
z(s) &=& \frac{\pi \sqrt{s} u(s)}{8}. \nn 
\eea

Note that several terms survive even if we take $m_l \rightarrow 0 $ 
limit. In that limit, the terms $T_5,T_6$ and $T_7$ of
Eq. (\ref{FuncT}) in $P_T^\pm$,
related to the coefficients of the scalar-- and tensor--type operators, 
still remain. Similarly, the terms $L_1, L_2, L_3, L_4$
and $L_8$ of Eq. (\ref{FuncL}) survive in the massless lepton limit.
Therefore, if we can measure the experimental values of 
$P^\pm_{T,L}$ for $B\rightarrow X_s \mu^+ \mu^-$, 
they may be very useful for searching for new interactions. 
Concerning the asymmetries $P_N^\pm $, 
while $P_N^- = - P_N^+$ in the SM,
the difference between $P_N^-$ and $-P_N^+$ comes only from
new interactions of the scalar- and tensor--type. By using this difference,
we may be able to study the information about the imaginary part of these
interactions. Of course, for  the massive lepton case
we can consider 
the difference between the asymmetries
$\tau^+ $ and $\tau^- $ to find the contribution of
new interactions. We are going to discuss this in the next Section. 

\section{Combined Analysis of the Asymmetries for $\tau^- $ and $\tau^+ $ } 

We can get very useful information to constrain the parameters and 
to find the evidence of new physics by measuring the asymmetries for 
each lepton, $\tau^+ $and $\tau^- $, and combining those asymmetries. 
We show  $P_L^- + P_L^+$, $P_T^- - P_T^+$,  and $P_N^- + P_N^+$, because
within the SM, $P_L^- + P_L^+ = 0$, $P_T^- - P_T^+ \approx 0$ and
$P_N^- + P_N^+ = 0$. 

({\bf A}) For $P_L^- + P_L^+$, the result is;
\bea
P_L^+ + P_L^- = \frac{{\cal B}_0~ u(s) }{m_b^8~ \frac{d {\cal B}}{d s}}
    &{\rm Re}[&  2 L_4(s)~\{ - C_{LRLR} C_{RLLR}^* 
                          + C_{LRRL} C_{RLRL}^* \} \nn \\
             &+& 4 L_5(s)~\{ m_b C_{BR} ( - C_{T}^* + 2 C_{TE}^* ) 
                             + m_s C_{SL}( C_{T}^* + 2 C_{TE}^* )\} \nn \\ 
             &+& 2 L_6(s)~ m_b\{ 2 (C_{LL} - C_{LR})
                                   ( C_{LRLR}^* + C_{LRRL}^* ) \nn \\
             & & ~~~~ + 2 (C_{RL} - C_{RR})
                                   ( C_{RLLR}^* + C_{RLRL}^* ) \nn \\
             & & ~~~~ - 8 ( C_{LL} + C_{LR}) ( C_T^* - 2
                                                  C_{TE}^* )
                                          \nn \\
             & & ~~~~  + 8 ( C_{RL} + C_{RR} ) ( C_T^* + 
                                           2 C_{TE}^* )    
                                                            \} \nn \\
             &+& 2 L_7(s)~ m_s\{ 2 (C_{LL} - C_{LR})
                                   ( C_{RLLR}^* + C_{RLRL}^* ) \nn \\
             & & ~~~~ + 2 (C_{RL} - C_{RR})
                                   ( C_{LRLR}^* + C_{LRRL}^* ) \nn \\
             & & ~~~~ - 8 ( C_{LL}+ C_{LR} ) ( C_T^* + 
                                             2 C_{TE}^* ) \nn \\
             & & ~~~~ + 8 ( C_{RL}+ C_{RR} ) ( C_T^* + 
                                             2 C_{TE}^* ) 
                                                    \}  \label{PLPL} \\
             &+& 2 L_8(s)~\{ - |C_{LRLR}|^2 + |C_{LRRL}|^2 
                            - |C_{RLLR}|^2 + |C_{RLRL}|^2 \} \nn \\
             &+& 2 (L_3(s) + L_8(s))~\{ - 128 C_T C_{TE}^*
                                                            \}]. \nn 
\eea
As can be seen from Eq. (21), 
the contribution from the SM to $P_L^+ + P_L^-$ completely disappears. 
There must be new interactions and new physics in
the decay $\Bstt$,
if the value of $P_L^+ + P_L^-$ is nonzero. 
We can then determine the parameters of
the scalar-- and tensor--type interactions, as shown 
in Eq. (\ref{PLPL}).
If we neglect the small strange quark mass $m_s$ ({\it i.e.} 
$ L_4$ and $L_7 $) and retain only terms with $m_b$,
then we find the following:
\begin{itemize}
\item There is no contribution from the combination of 
vector--type and nonlocal
operators because such terms are all canceled. 
\item 
The contribution from
$C_{LRLR} + C_{LRRL}$ is much
larger than from the other scalar--type coefficients, and 
it is
\bea 
&-& 8 m_b~ L_6(s)~ {\rm Re}[ C_{10} ( C_{LRLR}^* + C_{LRRL}^* ) ] \nn \\
&&~~ -~2 L_8(s)~ {\rm Re}[ ( C_{LRLR} + C_{LRRL} )( C_{LRLR}^* - C_{LRRL}^*
)]. 
\eea 
\item 
The tensor--type coefficients give the contribution
\bea
 &-& 256 ( L_3(s) + L_8(s) )~ {\rm Re}[C_{T} C_{TE}^*] \nn \\
&&~~ -~ 8 L_5(s)~ m_b C_{7}~ {\rm Re}( - C_{T}^* + 2 C_{TE}^* ) 
- 32 L_6(s)~ m_b~ {\rm Re}[ C_{9}^{\rm eff} ( C_T^* - 2 C_{TE}^* )]. 
\eea
\end{itemize}

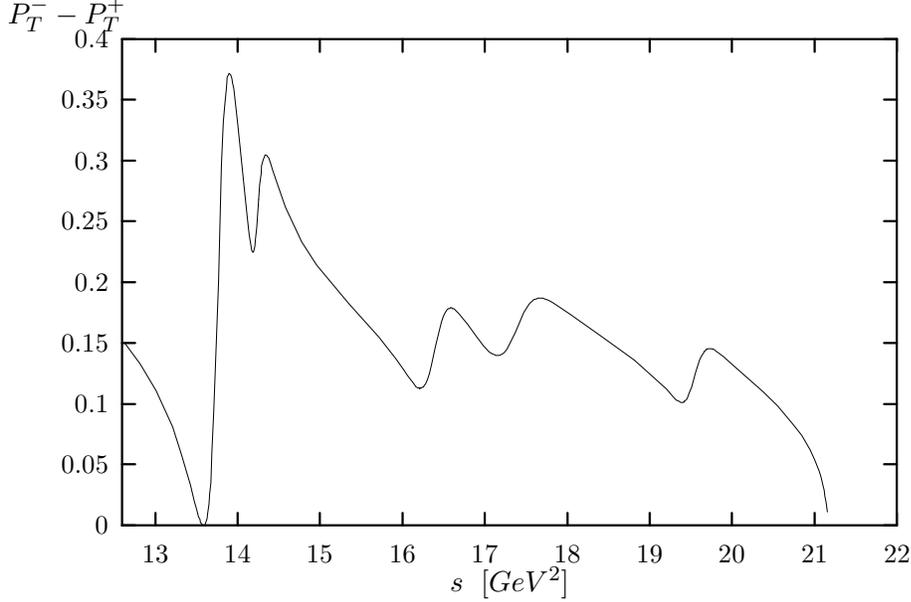
\begin{figure}[h]
\setlength{\unitlength}{0.240900pt}
\begin{picture}(1500,900)(0,0)
\tenrm
\thicklines \path(220,113)(240,113)
\thicklines \path(1436,113)(1416,113)
\put(198,113){\makebox(0,0)[r]{0}}
\thicklines \path(220,209)(240,209)
\thicklines \path(1436,209)(1416,209)
\put(198,209){\makebox(0,0)[r]{0.05}}
\thicklines \path(220,304)(240,304)
\thicklines \path(1436,304)(1416,304)
\put(198,304){\makebox(0,0)[r]{0.1}}
\thicklines \path(220,400)(240,400)
\thicklines \path(1436,400)(1416,400)
\put(198,400){\makebox(0,0)[r]{0.15}}
\thicklines \path(220,495)(240,495)
\thicklines \path(1436,495)(1416,495)
\put(198,495){\makebox(0,0)[r]{0.2}}
\thicklines \path(220,591)(240,591)
\thicklines \path(1436,591)(1416,591)
\put(198,591){\makebox(0,0)[r]{0.25}}
\thicklines \path(220,686)(240,686)
\thicklines \path(1436,686)(1416,686)
\put(198,686){\makebox(0,0)[r]{0.3}}
\thicklines \path(220,782)(240,782)
\thicklines \path(1436,782)(1416,782)
\put(198,782){\makebox(0,0)[r]{0.35}}
\thicklines \path(220,877)(240,877)
\thicklines \path(1436,877)(1416,877)
\put(198,877){\makebox(0,0)[r]{0.4}}
\thicklines \path(272,113)(272,133)
\thicklines \path(272,877)(272,857)
\put(272,68){\makebox(0,0){13}}
\thicklines \path(401,113)(401,133)
\thicklines \path(401,877)(401,857)
\put(401,68){\makebox(0,0){14}}
\thicklines \path(530,113)(530,133)
\thicklines \path(530,877)(530,857)
\put(530,68){\makebox(0,0){15}}
\thicklines \path(660,113)(660,133)
\thicklines \path(660,877)(660,857)
\put(660,68){\makebox(0,0){16}}
\thicklines \path(789,113)(789,133)
\thicklines \path(789,877)(789,857)
\put(789,68){\makebox(0,0){17}}
\thicklines \path(919,113)(919,133)
\thicklines \path(919,877)(919,857)
\put(919,68){\makebox(0,0){18}}
\thicklines \path(1048,113)(1048,133)
\thicklines \path(1048,877)(1048,857)
\put(1048,68){\makebox(0,0){19}}
\thicklines \path(1177,113)(1177,133)
\thicklines \path(1177,877)(1177,857)
\put(1177,68){\makebox(0,0){20}}
\thicklines \path(1307,113)(1307,133)
\thicklines \path(1307,877)(1307,857)
\put(1307,68){\makebox(0,0){21}}
\thicklines \path(1436,113)(1436,133)
\thicklines \path(1436,877)(1436,857)
\put(1436,68){\makebox(0,0){22}}
\thicklines \path(220,113)(1436,113)(1436,877)(220,877)(220,113)
\put(40,915){\makebox(0,0)[l]{\shortstack{$ P_T^- - P_T^+  $}}}
\put(828,23){\makebox(0,0){$ s ~~[GeV^2]$}}
\thinlines \path(224,399)(224,399)(248,367)(273,325)(299,268)(313,225)(327,176)(333,151)(337,139)(340,127)(342,122)(343,118)(344,116)(345,115)(347,113)(348,113)(350,115)(351,118)(353,124)(354,133)(356,146)(357,162)(359,180)(361,227)(364,293)(371,495)(376,679)(379,747)(381,776)(383,796)(384,808)(385,817)(387,822)(388,823)(390,821)(392,815)(393,808)(395,799)(401,745)(413,627)(416,598)(419,571)(421,561)(422,552)(423,545)(425,542)(426,543)(428,551)(429,563)(431,581)(434,625)
\thinlines \path(434,625)(436,648)(438,666)(439,678)(441,687)(443,692)(444,695)(445,695)(447,694)(448,693)(450,690)(453,681)(456,671)(463,652)(476,613)(502,559)(525,522)(575,462)(623,409)(650,374)(662,357)(668,348)(675,339)(678,335)(681,331)(682,330)(684,329)(686,329)(687,328)(688,329)(690,329)(691,330)(693,331)(694,333)(696,335)(699,342)(702,351)(705,364)(712,396)(718,421)(721,433)(724,442)(727,448)(729,451)(730,452)(732,454)(733,454)(734,454)(735,455)(737,455)(738,454)
\thinlines \path(738,454)(740,454)(741,453)(742,452)(748,446)(762,429)(775,411)(787,395)(794,388)(797,385)(800,383)(803,382)(805,381)(806,380)(808,380)(809,380)(809,380)(811,380)(812,380)(814,381)(815,382)(818,383)(821,386)(824,390)(837,416)(844,433)(850,447)(856,457)(859,461)(862,464)(864,466)(865,467)(867,468)(869,469)(870,469)(872,470)(873,470)(875,470)(876,470)(877,470)(878,470)(880,470)(881,470)(883,469)(884,469)(888,467)(894,464)(901,460)(924,444)(972,410)(1024,373)
\thinlines \path(1024,373)(1074,328)(1087,313)(1090,310)(1092,309)(1094,308)(1096,307)(1097,306)(1098,306)(1099,306)(1100,306)(1101,307)(1102,307)(1102,308)(1104,309)(1106,311)(1107,313)(1110,321)(1114,331)(1120,353)(1126,372)(1129,379)(1131,382)(1133,385)(1134,387)(1136,388)(1137,390)(1138,390)(1140,391)(1142,391)(1143,391)(1145,391)(1146,390)(1148,390)(1151,388)(1164,378)(1175,368)(1226,323)(1249,301)(1274,271)(1287,254)(1300,232)(1307,216)(1314,199)(1317,189)(1320,176)(1322,168)(1323,159)(1325,149)(1327,134)
\end{picture}
\caption{$P_T^- - P_T^+ $ in the SM.  } 
\label{PT-PT}
\end{figure}

({\bf B}) We now consider $P_T^- - P_T^+ $:
\bea
P_T^- - P_T^+  = \frac{ {\cal B}_0~ u(s) w(s)}
                       { m_b^8~ \frac{d {\cal B}}{d s}}&{\rm Re}[&
             4 m_l~\{ 2 m_b^2 C_{BR} ( C_{LL}^* - C_{LR}^* ) 
                            - 2  m_s^2 C_{SL} (C_{LL}^* - C_{LR}^* ) 
                                                  \nn \\ 
             & & ~~~~ - 2 m_b m_s C_{BR} ( C_{RL}^* - C_{RR}^* )
                               + 2 m_b m_s C_{SL} ( C_{RL}^* -  C_{RR}^* )
                            \} \nn \\
             &+& 4 T_3(s)~\{ 2 ( C_{T} - 2 C_{TE} ) 
                                     ( C_{LRLR}^* - C_{LRRL}^*) \nn \\
             & & ~~~~ + 2 ( C_{T} + 2 C_{TE} )
                                        (C_{RLLR}^* - C_{RLRL}^* )
                                               \} \nn \\
             &+& 2 T_5(s)~\{ 2 m_b C_{BR} ( - C_{LRLR}^* + C_{LRRL}^* ) \nn \\
             & & ~~~~ + 2 m_s C_{SL} ( - C_{RLLR}^* + C_{RLRL}^* )
                                                            \} \nn \\
             &+& 4 m_l^2~\{ m_b ( C_{LL} + C_{LR})(C_{LRLR}^* -
                                 C_{LRRL}^*) \nn \\
             & & ~~~~ +  m_s ( C_{LL} + C_{LR})(C_{RLLR}^* - C_{RLRL}^*) 
                                  \nn \\
             & & ~~~~ + m_s ( C_{RL} +C_{RR})(C_{LRRL}^* -
                                   C_{LRLR}^*) \nn \\
             & & ~~~~ +   m_b ( C_{RL} +C_{RR})(C_{RLLR}^* - C_{RLRL}^*)
                                                            \nn \\ 
             & & ~~~~ - 12 m_b ( C_{T} - 2 C_{TE}) 
                       ( C_{LL}^* - C_{RR}^* - C_{LR}^* + C_{RL}^*) \nn \\
             & & ~~~~ - 12 m_s (C_{T} + 2 C_{TE})
                                ( - C_{LL}^* + C_{RR}^*+ C_{LR}^* - C_{RL}^* )
                                                      \} \nn \\
             &+&  2 T_7(s)~\{ - m_b ( C_{LL} + C_{LR})(C_{LRLR}^* -
                                  C_{LRRL}^*) \nn \\
             & & ~~~~ -  m_s ( C_{LL} + C_{LR})(C_{RLLR}^* - C_{RLRL}^*) 
                                  \nn \\
             & & ~~~~ - m_s ( C_{RL} + C_{RR})(C_{LRLR}^* - C_{LRRL}^*)
                                   \nn \\
             & & ~~~~ -  m_b ( C_{RL} +C_{RR})(C_{RLLR}^* - C_{RLRL}^*) 
                                                            \nn \\ 
             & & ~~~~ + 4 m_s ( C_{T} + 2 C_{TE} )
                              ( C_{LL}^* - C_{RR}^* - C_{LR}^* + C_{RL}^* ) 
                                 \nn \\
             & & ~~~~ +   4 m_b ( C_{T} - 2 C_{TE} )
                           ( - C_{LL}^* + C_{RR}^*+ C_{LR}^* - C_{RL}^* ) 
                          \} ].
\label{PTPT}
\eea
The SM model contribution to $P_T^- - P_T^+ $ is only 
\bea
             4 m_l~\{ 2 m_b^2~ {\rm Re}[ C_{BR} ( C_{LL}^* &-& C_{LR}^* )] 
                       - 2  m_s^2~ {\rm Re}[ C_{SL} 
             (C_{LL}^* - C_{LR}^* )] \}   \nn \\
       &\simeq& 32 m_l~ [ m_b^2~ C_{7} C_{10}  
                            -   m_s^2~ C_{7} C_{10} ].  
                                                  \nn 
\label{PTPTinSM}
\eea 
We show in Fig. \ref{PT-PT}
the values of $P_T^- - P_T^+$ in the SM.
If we observe the experimental results different from Fig. 2,
we will have a strong evidence of the existence of physics beyond the SM.

After we neglect the terms with $m_s$, we get 
\bea
P_T^- - P_T^+  = \frac{ {\cal B}_0~ u(s) w(s)}
                       {m_b^8~ \frac{d {\cal B}}{d s}}&{\rm Re}[& 
             4 m_l~\{ 2 m_b^2 C_{BR} ( C_{LL}^* - C_{LR}^* ) 
                                                  \nn \\ 
             &+& 4 T_3(s)~\{ 2 ( C_{T} - 2 C_{TE} ) 
                                     ( C_{LRLR}^* - C_{LRRL}^*) \nn \\
             & & ~~~~ + 2 ( C_{T} + 2 C_{TE} )
                                        ( C_{RLLR}^* - C_{RLRL}^* )
                                               \} \nn \\
             &+& 2 T_5(s)~\{ 2 m_b C_{BR} ( - C_{LRLR}^* + C_{LRRL}^* )
                                                            \} \nn \\
             &+& 4 m_l^2~\{ m_b ( C_{LL} + C_{LR})(C_{LRLR}^* -
                                 C_{LRRL}^*) \nn \\
             & & ~~~~ +   m_b ( C_{RL} +C_{RR})(C_{RLLR}^* - C_{RLRL}^*)
                                                            \nn \\ 
             & & ~~~~ - 12 m_b ( C_{T} - 2 C_{TE}) 
                           ( C_{LL}^* - C_{RR}^* - C_{LR}^* + C_{RL}^*) 
                                                      \} \nn \\
             &+&  2 T_7(s)~\{ - m_b ( C_{LL} + C_{LR})(C_{LRLR}^* -
                                  C_{LRRL}^*) \nn \\
             & & ~~~~ -  m_b ( C_{RL} +C_{RR})(C_{RLLR}^* - C_{RLRL}^*) 
                                                            \nn \\ 
             & & ~~~~ +   4 m_b ( C_{T} - 2 C_{TE} )
                       ( - C_{LL}^* + C_{RR}^* + C_{LR}^* - C_{RL}^* ) 
                          \} ].
\label{PTPTnoms}
\eea
We find the following from the above expression:
\begin{itemize}
\item If there are only new vector--type interactions in addition to
the SM operators, the extra contributions come from 
\bea
- 16 m_l m_b^2~ C_7~ {\rm Re}( C^{*{\rm new}}_{LL} - C^{*{\rm new}}_{LR} ). 
\eea
Since 
$d{\cal B}/ds$ depends on $C_{LL}$ much more strongly than on 
any other Wilson coefficient
(this was found out in the previous work \cite{FKMY}),
$P_T^- - P_T^+$ is most sensitive to $C^{*{\rm new}}_{LL}$. 
\item  
The dependence on the scalar--type interactions is 
\bea
& & 4 m_b~ {\rm Re}[ \{ 2 T_5(s) C_{7} + ( 2 m_l^2 - T_7(s) ) 
      C_{9}^{\rm eff} \}  ( C_{LRLR}^* -  C_{LRRL}^* ) ] \nn \\
&=& 4 m_b s~ {\rm Re}[ \{ 2 C_{7} + C_{9}^{\rm eff} \} 
                                  ( C_{LRLR}^* -  C_{LRRL}^* ) ].
\eea
We find that here the contribution comes only from 
$(C_{LRLR}^* -  C_{LRRL}^*)$, 
in contrast to
the case of $P_L^- + P_L^+$. 
\item 
The dependence on the tensor--type interactions is
\bea
( 96 m_l^2 + 16 T_7(s) ) m_b~ C_{10} ( C_{T}^* - 2 C_{TE}^*) . 
\eea 
\end{itemize} 

({\bf C}) For $P_N^- + P_N^+$, we get
\bea
P_N^- + P_N^+  = - \frac{ {\cal B}_0~ u(s) z(s)}
                       {m_b^8~ \frac{d {\cal B}}{d s}}&{\rm Im}[& 
            4 ~\{ 
                  - m_b (C_{LL} - C_{LR})( C_{LRLR}^* - C_{LRRL}^* ) 
                            \nn \\
            & & ~~  - m_s (C_{LL} - C_{LR}) (C_{RLLR}^* - C_{RLRL}^* )
                                               \nn \\
            & & ~~ - m_s (C_{RL} - C_{RR})( C_{LRLR}^* - C_{LRRL}^* )
                                               \nn \\
            & & ~~ - m_b (C_{RL} - C_{RR})( C_{RLLR}^* - C_{RLRL}^* )
                                   \nn \\
            & & ~~ + 4 ( C_{T} -  2 C_{TE} )
                                    ( m_b ( C_{LL}^* + C_{LR}^* )
                                    - m_s ( C_{RL}^* + C_{RR}^* ) )
                                \label{PNPN} \\
            & & ~~ + 4 ( C_{T} + 2 C_{TE} )
                                    ( - m_s ( C_{LL}^* + C_{LR}^* )
                                      + m_b ( C_{RL}^* + C_{RR}^* ) 
                                          \}     \nn \\
             &+& \frac{32}{s} (m_b^2 - m_s^2)
                ~\{ m_b C_{BR} ( C_{T}^* -2 C_{TE}^* )
                            + m_s C_{SL} ( C_{T}^* +2 C_{TE}^* ) \}
            ]. \nn 
\eea
It appears that the experimental observation of the above quantity is much
more difficult than the other asymmetries because of the small numerical value.
However, if we measure it, 
we will be able to get information
about the imaginary parts of the coefficients 
for the scalar-- and tensor--type operators 
from Eq. (\ref{PNPN}). After we neglect 
small strange quark mass $m_s \to 0$, we find:
\begin{itemize}
\item 
The dependence on the scalar--type interactions is
\bea
- {\rm Im}[ 4 m_b~ (C_{LL} - C_{LR})( C_{LRLR}^* - C_{LRRL}^* ) ] = 
  8 m_b~ C_{10}~ {\rm Im}( C_{LRLR}^* - C_{LRRL}^* ). 
\eea
Therefore, we can get the  imaginary part of $C_{LRLR} - C_{LRRL} $
from the well known SM value of $C_{10}$. 
\item 
The dependence on the tensor--type interactions is
\bea
{\rm Im}[ 16 m_b~ ( C_{T}& -&  2 C_{TE} )( C_{LL}^* + C_{LR}^* )
- \frac{32}{s} m_b^2
                ~[ m_b C_{BR}^* ( C_{T} -2 C_{TE} )] \nn \\
&=& 32 m_b~ {\rm Im}[ ( 2 \frac{m_b^2}{s}C_7 + C_9^{*{\rm eff}} ) 
  ( C_{T} -2 C_{TE} ) ]. 
\eea
It would be quite difficult to 
determine (ascertain)
the imaginary parts of 
$C_{T} -2 C_{TE}, $ because $C_9^{\rm eff} $ is also a complex number. 
However, we
could still get some hints of the new physics.
\end{itemize}
                          
\section{ Averaged Values of the Asymmetries}

It may be experimentally difficult to measure the polarization asymmetries
of each lepton 
for all $l^+ l^-$ center--of--mass energies $s$.
However, we may get easily the averaged polarization
asymmetries. So we define the following averaged polarization 
asymmetries, 
\bea
<P_x> \equiv \frac{\int_{14~{\rm GeV}^2}^{(m_b - m_s)^2 } P_x(s) 
    \frac{d{\cal  B}}{ds} ds }
  {\int_{14~{\rm GeV}^2}^{(m_b - m_s)^2 } \frac{d{\cal B}}{ds} ds }, 
\eea
where we integrated from 14 GeV$^2$ to avoid the resonance regions below
$\psi^\prime $.
Within the SM, the results are
\bea
< P_L^\mp > = \mp 0.43, ~~&&< P_T^-> = - 0.56,~~ 
{\rm  and}~~ < P_N^\mp > = \pm 0.05. \nn \\  &&<P_T^+> = - 0.73, \nn
\eea
Note that here we neglected the final
state Coulomb interaction effect of the leptons with the other charged
particles, since the effect has been estimated to be much smaller 
than the averaged values of the SM \cite{Kruger,GN}. 
The final state interaction effect in 
the lepton polarization of $K_L\rightarrow \pi^+\mu^-\bar{\nu} $ or 
$K^+\rightarrow \pi^+\mu^+\mu^- $ is estimated as
the order of $10^{-3}$ \cite{FSI}. 
In this paper, we follow Ref.  \cite{Kruger,GN} and
neglect the final state interaction effect, because it will be
difficult to measure such small effect in experiments and  
decide whether the measurement includes the effects from some new
physics when  $< P_x^- > + < P_x^+ > $ is non
zero but very small value, where the subindex $x$ is $L$ or $N$. 

We summarize the findings from the averaged asymmetries as follows:
\begin{itemize}
\item
Within the SM,  $< P_L^- > + < P_L^+ > = 0$. 
As we find from Eq. (\ref{PLPL}), 
if there is any contribution from new interactions 
other than nonlocal--type and vector--type operators, 
$< P_L^- > + < P_L^+ > $ can have a nonzero value. The contributions
from the vector--type operators cancel each other. The main
contributions come from  
the terms with $ C_{LRLR} + C_{LRRL} $, $C_T$ and $C_{TE} $. 
\item
Within the SM,  $< P_T^- > - < P_T^+ > \simeq 0.17$.
The change of the value for $< P_T^- > - < P_T^+ > $ from the SM prediction 
would come from $C_{LRLR} - C_{LRRL} $ and $C_T - 2 C_{TE} $, 
as we discussed earlier. Therefore, 
for the scalar--type operators we can find rather easily
which type of operators gives contributions,
by using the measured values of both $< P_L^- > + < P_L^+ > $ and 
$< P_T^- > - < P_T^+ > $. 
\item
$< P_N^\mp >$ could give us interesting
information on the phases of new physics parameters. 
Within the SM $< P_N^- > + < P_N^+ > = 0$, and the experimentally
measured value, if it is nonzero, will give us some hints of the new
physics. 
\item
In Figs. \ref{BvsPL+PL}-\ref{BvsPNPNforCT}, 
we show correlations 
between the integrated branching ratio ${\cal B}$ and 
the averaged lepton polarized asymmetries 
of $\tau^-$ and $\tau^+$, by 
varying
the interaction coefficients. 
In drawing Figs.~\ref{BvsPL+PL}-\ref{BvsPNPNforCT},
we assume for simplicity that all the new interaction coefficients 
in Eq.(\ref{matrixelement}) are real. 
Since $< P_N^- > + < P_N^+ >$ comes only 
from the imaginary parts of the old (SM) coefficients, 
due to our temporary assumption of real new coefficients, 
the 
possible correlation 
between this combined asymmetry and the integrated branching ratio 
comes primarily ($m_s \approx 0$) from varying
$C_{T} - 2C_{TE}$, as seen from Eq. (31). 
\item
In Figs.\ref{BvsPL+PL}-\ref{BvsPLPLforCT},
we show the flows in $({\cal B}, {<P_L^-> + <P_L^+ > })$ plane,
where
\bea
 {\cal B} \equiv \int_{14~{\rm GeV}^2}^{(m_b - m_s)^2 } 
     \frac{d{\cal B}}{ds} ds, \nn
\eea 
by 
varying
the values of scalar--type coefficients (Fig.\ref{BvsPL+PL}),
and those of tensor--type coefficients (Fig.\ref{BvsPLPLforCT}).
In Figs. \ref{BvsPTPTforCLL}-\ref{BvsPTPTforCT}, we show the flows in
$({\cal B}, {<P_T^-> - <P_T^+ > })$ plane 
by 
varying
the values of vector--type (Fig.\ref{BvsPTPTforCLL}),
scalar--type (Fig.\ref{BvsPTPTforCLRLR})
and tensor--type coefficients (Fig.\ref{BvsPTPTforCT}). In
Fig.\ref{BvsPNPNforCT}, we show the flows in  
$({\cal B}, {<P_N^-> + <P_N^+ > })$ plane by changing the values of
tensor--type coefficients. 
\item
We note that the 
influence of varying 
of the coefficients in the Figures is
confirming our previous findings.
In Figs.\ref{BvsPL+PL}-\ref{BvsPLPLforCT}, we find 
that the influence of varying
$C_{LRLR}+C_{LRRL}$ and $C_T - 2 C_{TE}$ is quite
large. In Fig.\ref{BvsPTPTforCLL}, the flow by 
varying
$C_{LL}$ is
large.
In Figs.\ref{BvsPTPTforCLRLR}-\ref{BvsPTPTforCT}, 
we find 
that the variation of
$C_{LRLR}-C_{LRRL}$ 
and $C_T - 2 C_T$ will make large flow in the 
$({\cal B}, <P_T^-> - <P_T^+ > )$ plane. In Fig.\ref{BvsPNPNforCT},
we find 
that varying
$C_T - 2 C_{TE}$ will lead to a flow in the  
$({\cal B}, <P_N^-> + <P_N^+ > )$ plane, but it will not be so large one. 
\end{itemize}

\vspace{7cm}

\begin{figure}[hp]
\setlength{\unitlength}{0.240900pt}
\begin{picture}(1500,900)(0,0)
\tenrm
\thicklines \path(220,177)(240,177)
\thicklines \path(1436,177)(1416,177)
\put(198,177){\makebox(0,0)[r]{-1}}
\thicklines \path(220,336)(240,336)
\thicklines \path(1436,336)(1416,336)
\put(198,336){\makebox(0,0)[r]{-0.5}}
\thicklines \path(220,495)(240,495)
\thicklines \path(1436,495)(1416,495)
\put(198,495){\makebox(0,0)[r]{0}}
\thicklines \path(220,654)(240,654)
\thicklines \path(1436,654)(1416,654)
\put(198,654){\makebox(0,0)[r]{0.5}}
\thicklines \path(220,813)(240,813)
\thicklines \path(1436,813)(1416,813)
\put(198,813){\makebox(0,0)[r]{1}}
\thicklines \path(220,113)(220,133)
\thicklines \path(220,877)(220,857)
\put(220,68){\makebox(0,0){0}}
\thicklines \path(463,113)(463,133)
\thicklines \path(463,877)(463,857)
\put(463,68){\makebox(0,0){1}}
\thicklines \path(706,113)(706,133)
\thicklines \path(706,877)(706,857)
\put(706,68){\makebox(0,0){2}}
\thicklines \path(950,113)(950,133)
\thicklines \path(950,877)(950,857)
\put(950,68){\makebox(0,0){3}}
\thicklines \path(1193,113)(1193,133)
\thicklines \path(1193,877)(1193,857)
\put(1193,68){\makebox(0,0){4}}
\thicklines \path(1436,113)(1436,133)
\thicklines \path(1436,877)(1436,857)
\put(1436,68){\makebox(0,0){5}}
\thicklines \path(220,113)(1436,113)(1436,877)(220,877)(220,113)
\put(40,908){\makebox(0,0)[l]{\shortstack{$ < P_L^-> + < P_L^+>  $}}}
\put(828,23){\makebox(0,0){$ B \times 10^7  $}}
\Thicklines \path(1436,139)(1391,137)(1326,133)(1258,130)(1229,128)(1198,127)(1184,127)(1170,127)(1157,126)(1146,126)(1139,126)(1133,126)(1127,126)(1123,126)(1120,126)(1117,126)(1113,126)(1110,126)(1107,126)(1105,126)(1104,126)(1100,126)(1097,126)(1093,126)(1088,126)(1082,126)(1071,126)(1059,127)(1048,127)(1024,128)(1003,130)(981,132)(958,134)(939,137)(918,141)(884,149)(849,160)(816,175)(787,193)(764,212)(741,237)(722,263)(706,293)(691,329)(681,367)(676,386)(673,409)(670,430)(668,450)(667,461)(667,466)
\Thicklines \path(667,466)(667,473)(667,478)(666,484)(666,486)(666,489)(666,492)(666,494)(666,497)(666,500)(666,502)(666,504)(666,507)(666,510)(667,515)(667,525)(668,535)(668,546)(670,557)(672,578)(675,598)(680,619)(691,659)(704,693)(722,726)(743,756)(766,780)(793,801)(819,817)(849,830)(868,837)(885,842)(922,850)(944,854)(965,857)(986,859)(1009,861)(1029,862)(1052,863)(1064,863)(1071,864)(1077,864)(1083,864)(1088,864)(1094,864)(1097,864)(1100,864)(1103,864)(1105,864)(1107,864)(1110,864)
\Thicklines \path(1110,864)(1112,864)(1114,864)(1117,864)(1120,864)(1123,864)(1126,864)(1133,864)(1140,864)(1155,864)(1169,863)(1183,863)(1209,862)(1270,860)(1391,853)(1436,851)
\thinlines \path(1436,391)(1391,390)(1326,389)(1258,388)(1229,388)(1198,387)(1184,387)(1170,387)(1157,387)(1146,387)(1139,387)(1133,387)(1127,387)(1123,387)(1120,387)(1117,387)(1113,387)(1110,387)(1107,387)(1105,387)(1104,387)(1100,387)(1097,387)(1093,387)(1088,387)(1082,387)(1071,387)(1059,387)(1048,387)(1024,388)(1003,388)(981,389)(958,389)(939,390)(918,391)(884,394)(849,397)(816,401)(787,406)(764,412)(741,419)(722,427)(706,436)(691,446)(681,457)(676,463)(673,470)(670,476)(668,482)(667,485)(667,487)
\thinlines \path(667,487)(667,488)(667,490)(666,492)(666,492)(666,493)(666,494)(666,495)(666,496)(666,496)(666,497)(666,498)(666,499)(666,499)(667,501)(667,504)(668,507)(668,510)(670,513)(672,519)(675,525)(680,531)(691,543)(704,553)(722,563)(743,571)(766,579)(793,585)(819,589)(849,593)(868,595)(885,597)(922,599)(944,600)(965,601)(986,602)(1009,602)(1029,602)(1052,603)(1064,603)(1071,603)(1077,603)(1083,603)(1088,603)(1094,603)(1097,603)(1100,603)(1103,603)(1105,603)(1107,603)(1110,603)
\thinlines \path(1110,603)(1112,603)(1114,603)(1117,603)(1120,603)(1123,603)(1126,603)(1133,603)(1140,603)(1155,603)(1169,603)(1183,603)(1209,603)(1270,602)(1391,600)(1436,599)
\end{picture} 
\caption{ The flows in $({\cal B}, {<P_L^-> + <P_L^+ > })$ plane.
In each flow, $C_{LRLR } + C_{LRRL}$ (thick solid line), 
$C_{RLLR}  + C_{RLRL}$ (thin line)
are varied respectively. }
\label{BvsPL+PL}
\end{figure}
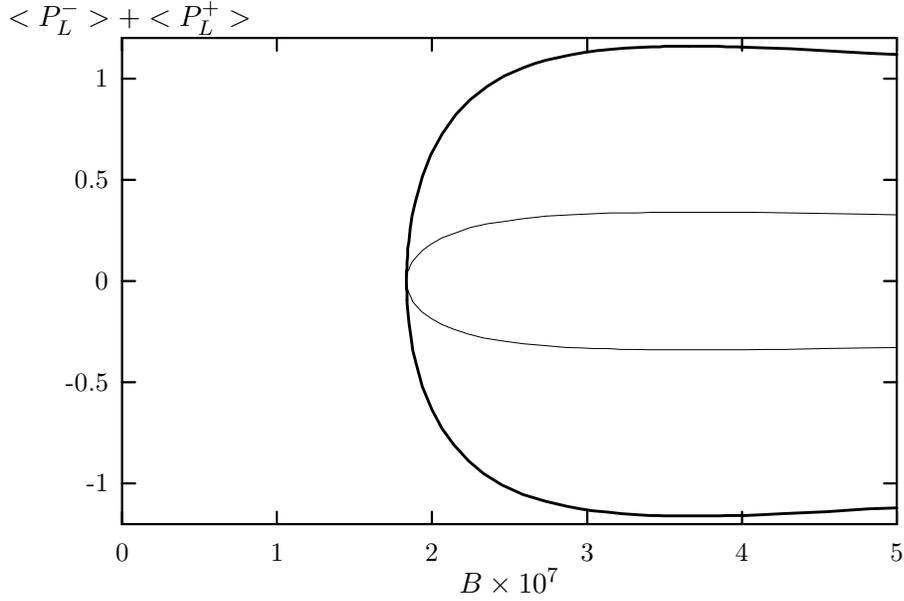

\begin{figure}[hp]
\setlength{\unitlength}{0.240900pt}
\begin{picture}(1500,900)(0,0)
\tenrm
\thicklines \path(220,113)(240,113)
\thicklines \path(1436,113)(1416,113)
\put(198,113){\makebox(0,0)[r]{-1}}
\thicklines \path(220,189)(240,189)
\thicklines \path(1436,189)(1416,189)
\put(198,189){\makebox(0,0)[r]{-0.8}}
\thicklines \path(220,266)(240,266)
\thicklines \path(1436,266)(1416,266)
\put(198,266){\makebox(0,0)[r]{-0.6}}
\thicklines \path(220,342)(240,342)
\thicklines \path(1436,342)(1416,342)
\put(198,342){\makebox(0,0)[r]{-0.4}}
\thicklines \path(220,419)(240,419)
\thicklines \path(1436,419)(1416,419)
\put(198,419){\makebox(0,0)[r]{-0.2}}
\thicklines \path(220,495)(240,495)
\thicklines \path(1436,495)(1416,495)
\put(198,495){\makebox(0,0)[r]{0}}
\thicklines \path(220,571)(240,571)
\thicklines \path(1436,571)(1416,571)
\put(198,571){\makebox(0,0)[r]{0.2}}
\thicklines \path(220,648)(240,648)
\thicklines \path(1436,648)(1416,648)
\put(198,648){\makebox(0,0)[r]{0.4}}
\thicklines \path(220,724)(240,724)
\thicklines \path(1436,724)(1416,724)
\put(198,724){\makebox(0,0)[r]{0.6}}
\thicklines \path(220,801)(240,801)
\thicklines \path(1436,801)(1416,801)
\put(198,801){\makebox(0,0)[r]{0.8}}
\thicklines \path(220,877)(240,877)
\thicklines \path(1436,877)(1416,877)
\put(198,877){\makebox(0,0)[r]{1}}
\thicklines \path(220,113)(220,133)
\thicklines \path(220,877)(220,857)
\put(220,68){\makebox(0,0){0}}
\thicklines \path(463,113)(463,133)
\thicklines \path(463,877)(463,857)
\put(463,68){\makebox(0,0){1}}
\thicklines \path(706,113)(706,133)
\thicklines \path(706,877)(706,857)
\put(706,68){\makebox(0,0){2}}
\thicklines \path(950,113)(950,133)
\thicklines \path(950,877)(950,857)
\put(950,68){\makebox(0,0){3}}
\thicklines \path(1193,113)(1193,133)
\thicklines \path(1193,877)(1193,857)
\put(1193,68){\makebox(0,0){4}}
\thicklines \path(1436,113)(1436,133)
\thicklines \path(1436,877)(1436,857)
\put(1436,68){\makebox(0,0){5}}
\thicklines \path(220,113)(1436,113)(1436,877)(220,877)(220,113)
\put(40,908){\makebox(0,0)[l]{\shortstack{$ < P_L^-> + < P_L^+>  $}}}
\put(828,23){\makebox(0,0){$ B \times 10^7  $}}
\Thicklines \path(1436,644)(1327,634)(1217,623)(1108,609)(939,580)(869,564)(806,546)(713,513)(682,501)(672,497)(664,494)(661,493)(660,493)(658,493)(657,492)(657,492)(656,492)(656,492)(656,492)(655,492)(655,492)(655,492)(655,492)(654,492)(654,493)(654,493)(654,493)(654,493)(654,493)(654,493)(654,494)(654,494)(654,494)(655,495)(655,496)(659,500)(666,505)(684,518)(749,553)(797,572)(859,592)(928,609)(1015,627)(1109,642)(1206,654)(1433,675)(1436,675)
\thinlines \path(1436,214)(1392,215)(1324,217)(1202,223)(1140,226)(1084,230)(982,239)(897,250)(819,264)(779,273)(744,283)(711,294)(681,306)(634,331)(612,346)(592,363)(563,394)(551,411)(541,429)(534,445)(532,451)(530,459)(529,466)(528,469)(528,471)(528,473)(528,474)(528,476)(527,477)(527,479)(527,480)(527,482)(528,483)(528,484)(528,486)(528,487)(529,490)(529,493)(530,495)(532,500)(534,502)(536,504)(539,507)(541,508)(543,510)(545,511)(548,511)(550,512)(552,513)(555,513)(557,513)
\thinlines \path(557,513)(559,514)(560,514)(561,514)(561,514)(562,514)(563,514)(564,514)(566,514)(568,514)(570,514)(571,514)(574,514)(577,514)(581,513)(590,512)(600,510)(619,506)(670,494)(792,465)(965,434)(1171,408)(1407,387)(1436,385)
\end{picture}
\caption{ The flows in $({\cal B}, {<P_L^-> + <P_L^+ > })$ plane.
In each flow, $C_T  + 2 C_{TE}$ (thick solid line), 
$C_T  - 2 C_{TE}$ (thin line)
are varied respectively. }
\label{BvsPLPLforCT}
\end{figure}
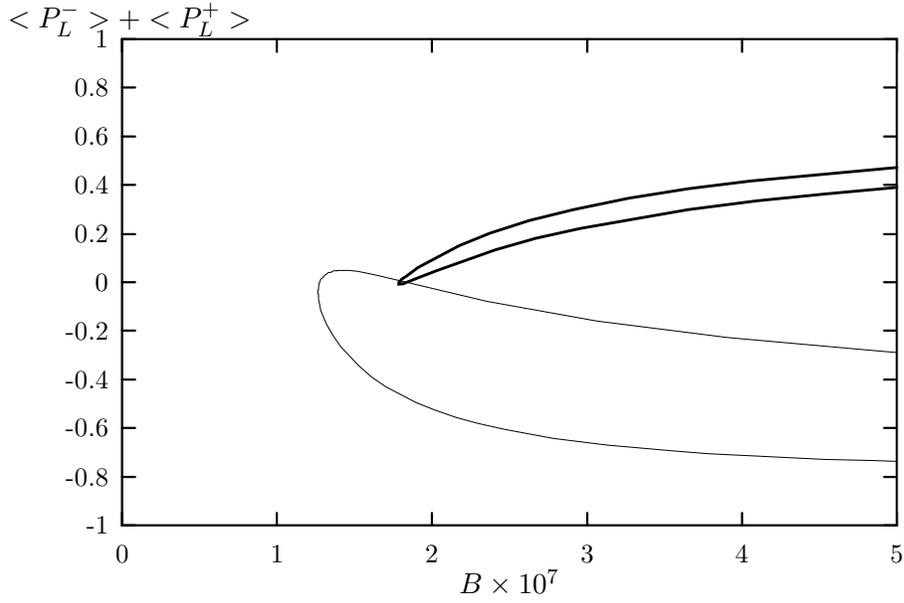

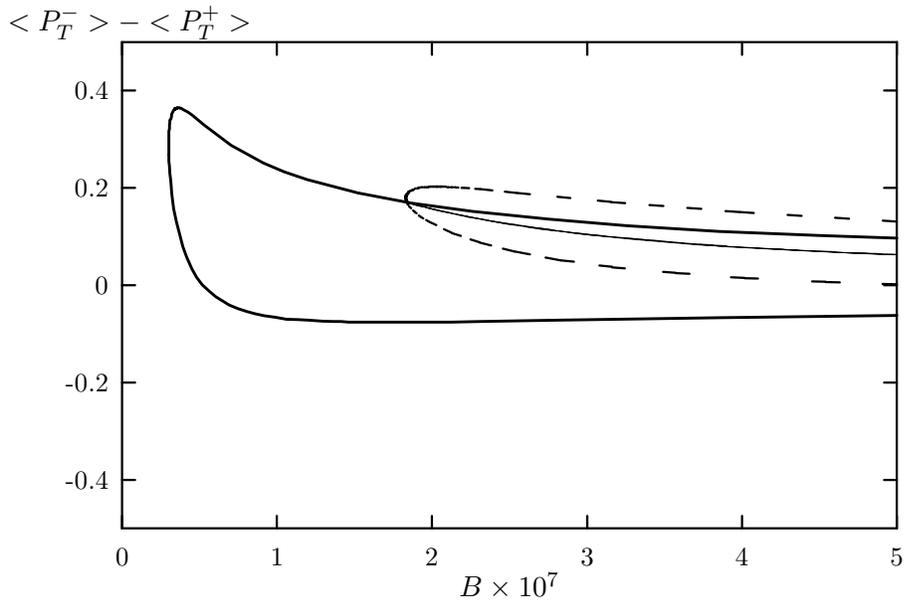
\begin{figure}[hp]
\setlength{\unitlength}{0.240900pt}
\begin{picture}(1500,900)(0,0)
\tenrm
\thicklines \path(220,189)(240,189)
\thicklines \path(1436,189)(1416,189)
\put(198,189){\makebox(0,0)[r]{-0.4}}
\thicklines \path(220,342)(240,342)
\thicklines \path(1436,342)(1416,342)
\put(198,342){\makebox(0,0)[r]{-0.2}}
\thicklines \path(220,495)(240,495)
\thicklines \path(1436,495)(1416,495)
\put(198,495){\makebox(0,0)[r]{0}}
\thicklines \path(220,648)(240,648)
\thicklines \path(1436,648)(1416,648)
\put(198,648){\makebox(0,0)[r]{0.2}}
\thicklines \path(220,801)(240,801)
\thicklines \path(1436,801)(1416,801)
\put(198,801){\makebox(0,0)[r]{0.4}}
\thicklines \path(220,113)(220,133)
\thicklines \path(220,877)(220,857)
\put(220,68){\makebox(0,0){0}}
\thicklines \path(463,113)(463,133)
\thicklines \path(463,877)(463,857)
\put(463,68){\makebox(0,0){1}}
\thicklines \path(706,113)(706,133)
\thicklines \path(706,877)(706,857)
\put(706,68){\makebox(0,0){2}}
\thicklines \path(950,113)(950,133)
\thicklines \path(950,877)(950,857)
\put(950,68){\makebox(0,0){3}}
\thicklines \path(1193,113)(1193,133)
\thicklines \path(1193,877)(1193,857)
\put(1193,68){\makebox(0,0){4}}
\thicklines \path(1436,113)(1436,133)
\thicklines \path(1436,877)(1436,857)
\put(1436,68){\makebox(0,0){5}}
\thicklines \path(220,113)(1436,113)(1436,877)(220,877)(220,113)
\put(40,908){\makebox(0,0)[l]{\shortstack{$ < P_T^-> - < P_T^+>  $}}}
\put(828,23){\makebox(0,0){$ B \times 10^7  $}}
\Thicklines \path(1436,448)(1160,444)(897,440)(782,438)(736,437)(712,437)(688,437)(676,437)(669,437)(663,437)(657,437)(651,437)(645,437)(640,437)(635,437)(629,437)(624,437)(620,437)(614,437)(608,437)(598,437)(587,437)(575,437)(563,438)(553,438)(534,439)(514,440)(495,441)(478,442)(461,445)(443,447)(429,450)(416,454)(402,458)(388,464)(367,477)(356,486)(346,496)(337,508)(329,521)(321,539)(315,556)(305,594)(301,615)(298,639)(296,659)(295,671)(294,681)(293,690)(293,695)(293,700)
\Thicklines \path(293,700)(293,705)(293,709)(293,713)(293,717)(293,721)(293,725)(293,729)(293,733)(293,737)(294,741)(294,747)(295,750)(295,754)(296,756)(297,759)(298,764)(299,767)(301,770)(302,771)(302,772)(303,772)(304,773)(306,773)(307,774)(308,774)(309,774)(310,774)(311,773)(312,773)(313,773)(316,771)(323,767)(330,762)(348,747)(391,715)(442,687)(475,673)(510,661)(590,640)(674,624)(770,611)(888,599)(1012,589)(1157,580)(1318,573)(1436,569)
\thicklines \dashline[-10]{25}(1436,595)(1189,610)(987,626)(839,640)(782,646)(756,648)(743,649)(737,649)(731,649)(727,650)(724,650)(722,650)(720,650)(719,650)(718,650)(716,650)(714,650)(713,650)(712,650)(710,650)(708,650)(706,650)(704,650)(702,650)(700,649)(697,649)(694,649)(690,649)(685,648)(683,647)(680,647)(676,645)(674,644)(672,643)(669,642)(668,640)(667,639)(666,638)(666,637)(665,637)(665,636)(665,635)(665,634)(664,634)(664,633)(664,632)(664,632)(664,631)(665,630)(665,630)(665,628)
\thicklines \dashline[-10]{25}(665,628)(666,627)(666,625)(668,622)(671,618)(679,610)(689,602)(701,595)(737,577)(781,562)(836,548)(909,534)(989,523)(1089,514)(1197,506)(1324,500)(1436,496)
\thinlines \dashline[-10]{25}(1436,543)(1381,545)(1256,551)(1136,558)(954,574)(877,583)(808,594)(710,614)(680,621)(664,626)(660,627)(659,627)(659,627)(659,627)(659,627)(659,627)(659,627)(659,627)(661,627)(665,626)(717,612)(762,602)(815,593)(956,574)(1041,566)(1147,558)(1258,551)(1390,545)(1436,543)
\thinlines \path(1436,543)(1423,543)(1293,549)(1169,556)(979,572)(898,581)(824,591)(717,612)(663,626)(657,628)(654,629)(654,629)(654,629)(654,629)(654,629)(654,629)(655,629)(656,628)(670,624)(698,617)(736,608)(791,597)(931,577)(1018,568)(1124,559)(1348,547)(1436,543)
\end{picture} 
\caption{ The flows in $({\cal B}, {<P_T^-> - <P_T^+ > })$ plane.
In each flow, $C_{LL}$ (thick solid line), 
$C_{LR}$ (thick dashed line), $C_{RL}$, $C_{RR}$ (thin solid line) 
are varied respectively. }
\label{BvsPTPTforCLL}
\end{figure}

\begin{figure}[hp]
\setlength{\unitlength}{0.240900pt}
\begin{picture}(1500,900)(0,0)
\tenrm
\thicklines \path(220,113)(240,113)
\thicklines \path(1436,113)(1416,113)
\put(198,113){\makebox(0,0)[r]{-1}}
\thicklines \path(220,240)(240,240)
\thicklines \path(1436,240)(1416,240)
\put(198,240){\makebox(0,0)[r]{-0.5}}
\thicklines \path(220,368)(240,368)
\thicklines \path(1436,368)(1416,368)
\put(198,368){\makebox(0,0)[r]{0}}
\thicklines \path(220,495)(240,495)
\thicklines \path(1436,495)(1416,495)
\put(198,495){\makebox(0,0)[r]{0.5}}
\thicklines \path(220,622)(240,622)
\thicklines \path(1436,622)(1416,622)
\put(198,622){\makebox(0,0)[r]{1}}
\thicklines \path(220,750)(240,750)
\thicklines \path(1436,750)(1416,750)
\put(198,750){\makebox(0,0)[r]{1.5}}
\thicklines \path(220,877)(240,877)
\thicklines \path(1436,877)(1416,877)
\put(198,877){\makebox(0,0)[r]{2}}
\thicklines \path(220,113)(220,133)
\thicklines \path(220,877)(220,857)
\put(220,68){\makebox(0,0){0}}
\thicklines \path(463,113)(463,133)
\thicklines \path(463,877)(463,857)
\put(463,68){\makebox(0,0){1}}
\thicklines \path(706,113)(706,133)
\thicklines \path(706,877)(706,857)
\put(706,68){\makebox(0,0){2}}
\thicklines \path(950,113)(950,133)
\thicklines \path(950,877)(950,857)
\put(950,68){\makebox(0,0){3}}
\thicklines \path(1193,113)(1193,133)
\thicklines \path(1193,877)(1193,857)
\put(1193,68){\makebox(0,0){4}}
\thicklines \path(1436,113)(1436,133)
\thicklines \path(1436,877)(1436,857)
\put(1436,68){\makebox(0,0){5}}
\thicklines \path(220,113)(1436,113)(1436,877)(220,877)(220,113)
\put(40,908){\makebox(0,0)[l]{\shortstack{$ < P_T^-> - < P_T^+>  $}}}
\put(828,23){\makebox(0,0){$ B \times 10^7  $}}
\Thicklines \path(1436,384)(1397,384)(1258,386)(1138,389)(1026,392)(932,395)(847,399)(781,402)(732,406)(694,408)(673,410)(668,411)(667,411)(666,411)(666,411)(666,411)(666,411)(666,411)(666,411)(666,411)(667,411)(668,411)(674,410)(696,408)(731,406)(783,402)(847,399)(931,395)(1027,392)(1133,389)(1262,386)(1400,384)(1436,384)
\thicklines \dashline[-10]{25}(1436,285)(1371,286)(1293,289)(1213,293)(1142,297)(1081,302)(1015,308)(958,315)(899,325)(841,338)(793,351)(746,368)(706,387)(672,407)(638,433)(610,460)(583,493)(559,533)(539,574)(531,596)(524,619)(518,640)(513,663)(510,683)(508,700)(508,705)(507,710)(507,714)(507,719)(507,721)(506,724)(506,728)(506,730)(506,733)(506,735)(506,737)(506,742)(507,746)(507,749)(507,753)(508,760)(509,767)(510,773)(511,778)(513,784)(515,789)(519,797)(522,802)(524,805)(530,811)(534,814)
\thicklines \dashline[-10]{25}(534,814)(538,816)(542,817)(546,819)(549,819)(551,820)(552,820)(554,820)(556,820)(559,820)(561,820)(564,820)(565,820)(567,820)(570,820)(572,820)(578,819)(584,818)(589,817)(600,814)(736,768)(837,736)(945,707)(1209,654)(1362,632)(1436,623)
\thinlines \path(1436,384)(1397,384)(1258,386)(1138,389)(1026,392)(932,395)(847,399)(781,402)(732,406)(694,408)(673,410)(668,411)(667,411)(666,411)(666,411)(666,411)(666,411)(666,411)(666,411)(666,411)(667,411)(668,411)(674,410)(696,408)(731,406)(783,402)(847,399)(931,395)(1027,392)(1133,389)(1262,386)(1400,384)(1436,384)
\thinlines \dashline[-10]{25}(1436,376)(1414,377)(1277,378)(1141,381)(1032,384)(934,387)(844,391)(775,395)(718,400)(697,402)(679,405)(667,407)(662,407)(658,408)(657,409)(656,409)(655,409)(655,409)(655,410)(654,410)(654,410)(654,410)(654,410)(654,410)(654,410)(654,410)(655,410)(655,410)(655,411)(655,411)(656,411)(656,411)(658,411)(658,411)(659,411)(660,411)(661,411)(661,411)(662,411)(663,411)(663,411)(664,411)(664,411)(666,411)(667,411)(668,411)(668,411)(671,411)(674,411)(681,411)(696,410)
\thinlines \dashline[-10]{25}(696,410)(742,408)(882,402)(1076,396)(1200,393)(1349,391)(1436,389)
\end{picture}
\caption{ The flows in $({\cal B}, {<P_T^-> - <P_T^+ > })$ plane.
In each flow, $C_{LRLR} + C_{LRRL}$ (thick solid line), 
$C_{LRLR} - C_{LRRL} $ (thick dashed line), $C_{RLLR} + C_{RLRL}$
(thck solid line), $C_{RLLR}- C_{RLRL}$ (thin dashed  line) 
are varied respectively. }
\label{BvsPTPTforCLRLR}
\end{figure}
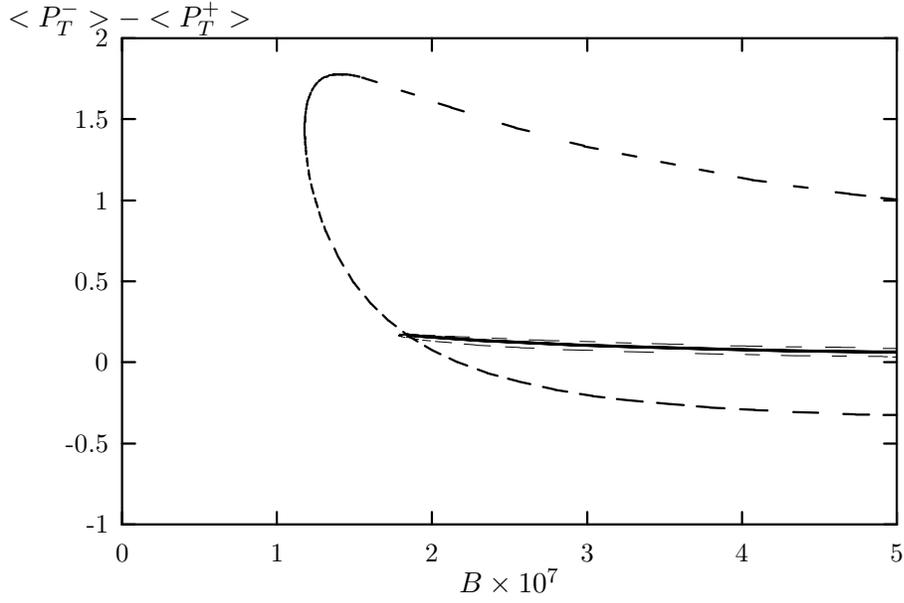

\begin{figure}[hp]
\setlength{\unitlength}{0.240900pt}
\begin{picture}(1500,900)(0,0)
\tenrm
\thicklines \path(220,113)(240,113)
\thicklines \path(1436,113)(1416,113)
\put(198,113){\makebox(0,0)[r]{-1}}
\thicklines \path(220,189)(240,189)
\thicklines \path(1436,189)(1416,189)
\put(198,189){\makebox(0,0)[r]{-0.8}}
\thicklines \path(220,266)(240,266)
\thicklines \path(1436,266)(1416,266)
\put(198,266){\makebox(0,0)[r]{-0.6}}
\thicklines \path(220,342)(240,342)
\thicklines \path(1436,342)(1416,342)
\put(198,342){\makebox(0,0)[r]{-0.4}}
\thicklines \path(220,419)(240,419)
\thicklines \path(1436,419)(1416,419)
\put(198,419){\makebox(0,0)[r]{-0.2}}
\thicklines \path(220,495)(240,495)
\thicklines \path(1436,495)(1416,495)
\put(198,495){\makebox(0,0)[r]{0}}
\thicklines \path(220,571)(240,571)
\thicklines \path(1436,571)(1416,571)
\put(198,571){\makebox(0,0)[r]{0.2}}
\thicklines \path(220,648)(240,648)
\thicklines \path(1436,648)(1416,648)
\put(198,648){\makebox(0,0)[r]{0.4}}
\thicklines \path(220,724)(240,724)
\thicklines \path(1436,724)(1416,724)
\put(198,724){\makebox(0,0)[r]{0.6}}
\thicklines \path(220,801)(240,801)
\thicklines \path(1436,801)(1416,801)
\put(198,801){\makebox(0,0)[r]{0.8}}
\thicklines \path(220,877)(240,877)
\thicklines \path(1436,877)(1416,877)
\put(198,877){\makebox(0,0)[r]{1}}
\thicklines \path(220,113)(220,133)
\thicklines \path(220,877)(220,857)
\put(220,68){\makebox(0,0){0}}
\thicklines \path(463,113)(463,133)
\thicklines \path(463,877)(463,857)
\put(463,68){\makebox(0,0){1}}
\thicklines \path(706,113)(706,133)
\thicklines \path(706,877)(706,857)
\put(706,68){\makebox(0,0){2}}
\thicklines \path(950,113)(950,133)
\thicklines \path(950,877)(950,857)
\put(950,68){\makebox(0,0){3}}
\thicklines \path(1193,113)(1193,133)
\thicklines \path(1193,877)(1193,857)
\put(1193,68){\makebox(0,0){4}}
\thicklines \path(1436,113)(1436,133)
\thicklines \path(1436,877)(1436,857)
\put(1436,68){\makebox(0,0){5}}
\thicklines \path(220,113)(1436,113)(1436,877)(220,877)(220,113)
\put(40,908){\makebox(0,0)[l]{\shortstack{$ < P_T^-> - < P_T^+>  $}}}
\put(828,23){\makebox(0,0){$ B \times 10^7  $}}
\Thicklines \path(1436,515)(1410,515)(1204,520)(1113,523)(1026,527)(892,535)(835,539)(783,544)(741,548)(705,553)(679,558)(664,561)(659,562)(657,562)(655,563)(655,563)(654,563)(654,563)(654,563)(654,563)(654,564)(654,564)(654,564)(654,564)(654,564)(654,564)(654,564)(654,564)(655,564)(655,564)(655,564)(655,564)(655,564)(656,564)(656,564)(656,564)(657,564)(657,564)(657,564)(658,564)(659,564)(660,564)(663,564)(678,562)(731,557)(821,549)(940,541)(1013,537)(1101,534)(1290,528)(1397,525)
\Thicklines \path(1397,525)(1436,524)
\thinlines \path(1436,689)(1336,698)(1123,724)(949,752)(797,785)(682,815)(636,826)(615,831)(605,833)(600,833)(595,834)(591,834)(588,835)(585,835)(583,835)(582,835)(580,835)(579,835)(578,835)(577,835)(575,835)(574,835)(573,835)(572,834)(568,834)(567,834)(565,834)(559,832)(556,831)(554,830)(549,828)(545,825)(541,822)(538,818)(535,813)(532,807)(530,802)(529,796)(529,793)(528,790)(528,788)(528,786)(528,784)(528,783)(527,781)(527,779)(527,777)(527,775)(527,773)(528,772)(528,771)
\thinlines \path(528,771)(528,767)(528,763)(529,758)(530,750)(532,740)(535,729)(541,709)(549,690)(560,669)(584,632)(601,613)(622,593)(666,561)(717,534)(746,522)(780,511)(847,494)(884,486)(927,479)(1013,468)(1064,463)(1113,459)(1218,453)(1321,448)(1435,445)(1436,445)
\end{picture}
\caption{ The flows in $({\cal B}, {<P_T^-> - <P_T^+ > })$ plane.
In each flow, $C_{T} + 2 C_{TE}$ (thick solid line), 
$C_{T} - 2 C_{TE} $ (thin solid line)
are varied respectively. }
\label{BvsPTPTforCT}
\end{figure}
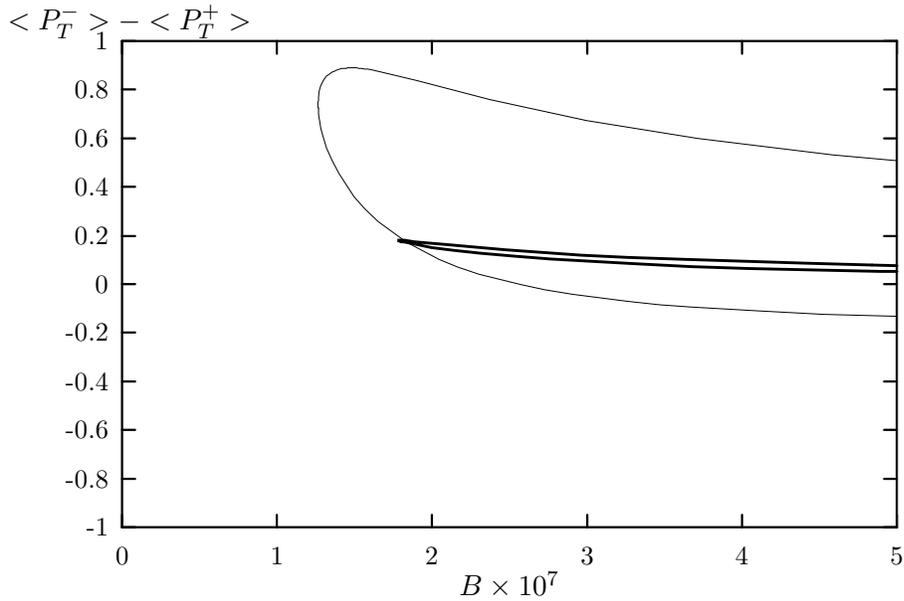

\begin{figure}[ht]
\setlength{\unitlength}{0.240900pt}
\begin{picture}(1500,900)(0,0)
\tenrm
\thicklines \path(220,113)(240,113)
\thicklines \path(1436,113)(1416,113)
\put(198,113){\makebox(0,0)[r]{-0.2}}
\thicklines \path(220,208)(240,208)
\thicklines \path(1436,208)(1416,208)
\put(198,208){\makebox(0,0)[r]{-0.15}}
\thicklines \path(220,304)(240,304)
\thicklines \path(1436,304)(1416,304)
\put(198,304){\makebox(0,0)[r]{-0.1}}
\thicklines \path(220,400)(240,400)
\thicklines \path(1436,400)(1416,400)
\put(198,400){\makebox(0,0)[r]{-0.05}}
\thicklines \path(220,495)(240,495)
\thicklines \path(1436,495)(1416,495)
\put(198,495){\makebox(0,0)[r]{0}}
\thicklines \path(220,591)(240,591)
\thicklines \path(1436,591)(1416,591)
\put(198,591){\makebox(0,0)[r]{0.05}}
\thicklines \path(220,686)(240,686)
\thicklines \path(1436,686)(1416,686)
\put(198,686){\makebox(0,0)[r]{0.1}}
\thicklines \path(220,782)(240,782)
\thicklines \path(1436,782)(1416,782)
\put(198,782){\makebox(0,0)[r]{0.15}}
\thicklines \path(220,877)(240,877)
\thicklines \path(1436,877)(1416,877)
\put(198,877){\makebox(0,0)[r]{0.2}}
\thicklines \path(220,113)(220,133)
\thicklines \path(220,877)(220,857)
\put(220,68){\makebox(0,0){0}}
\thicklines \path(463,113)(463,133)
\thicklines \path(463,877)(463,857)
\put(463,68){\makebox(0,0){1}}
\thicklines \path(706,113)(706,133)
\thicklines \path(706,877)(706,857)
\put(706,68){\makebox(0,0){2}}
\thicklines \path(950,113)(950,133)
\thicklines \path(950,877)(950,857)
\put(950,68){\makebox(0,0){3}}
\thicklines \path(1193,113)(1193,133)
\thicklines \path(1193,877)(1193,857)
\put(1193,68){\makebox(0,0){4}}
\thicklines \path(1436,113)(1436,133)
\thicklines \path(1436,877)(1436,857)
\put(1436,68){\makebox(0,0){5}}
\thicklines \path(220,113)(1436,113)(1436,877)(220,877)(220,113)
\put(40,908){\makebox(0,0)[l]{\shortstack{$ < P_N^-> + < P_N^+>  $}}}
\put(828,23){\makebox(0,0){$ B \times 10^7  $}}
\thinlines \path(1436,319)(1342,312)(1006,277)(859,256)(745,237)(698,230)(659,225)(641,223)(636,223)(631,223)(623,222)(619,222)(614,222)(607,222)(603,223)(599,223)(595,223)(591,224)(585,225)(579,226)(574,227)(568,229)(562,232)(557,234)(553,237)(549,240)(543,247)(538,254)(534,263)(532,268)(530,274)(529,280)(528,283)(528,287)(528,293)(527,300)(527,303)(528,307)(528,315)(529,319)(529,323)(531,332)(535,348)(539,362)(546,379)(554,394)(573,422)(598,450)(625,471)(659,491)(700,509)
\thinlines \path(700,509)(742,523)(794,535)(855,546)(912,553)(971,558)(1039,562)(1079,564)(1117,566)(1200,569)(1244,570)(1294,570)(1337,571)(1386,572)(1433,572)(1436,572)
\Thicklines \path(1436,499)(1418,499)(1367,499)(1345,499)(1322,499)(1310,499)(1297,499)(1285,499)(1274,499)(1254,499)(1235,499)(1214,499)(1195,499)(1177,499)(1157,499)(1137,499)(1119,499)(1087,499)(1054,499)(1019,499)(983,499)(955,499)(925,499)(897,499)(869,499)(845,499)(822,498)(782,498)(750,498)(734,497)(719,497)(695,496)(683,496)(674,495)(667,495)(664,495)(661,495)(659,494)(657,494)(656,494)(655,494)(655,494)(654,494)(654,493)(654,493)(654,493)(655,493)(655,493)(657,493)(658,492)(660,492)
\Thicklines \path(660,492)(664,492)(670,491)(677,491)(687,491)(700,490)(713,490)(741,490)(759,490)(777,489)(795,489)(816,489)(840,489)(852,489)(864,489)(875,489)(887,489)(898,489)(910,489)(923,489)(937,489)(951,489)(964,489)(979,489)(987,489)(995,489)(1012,489)(1030,489)(1063,489)(1095,489)(1263,489)(1436,489)
\end{picture}
\caption{ The flows in $({\cal B}, {<P_N^-> + <P_N^+> })$ plane.
In each flow, $C_{T} + 2 C_{TE}$ (thick solid line), 
$C_{T} - 2 C_{TE} $ (thin solid line)
are varied respectively. }
\label{BvsPNPNforCT}
\end{figure}
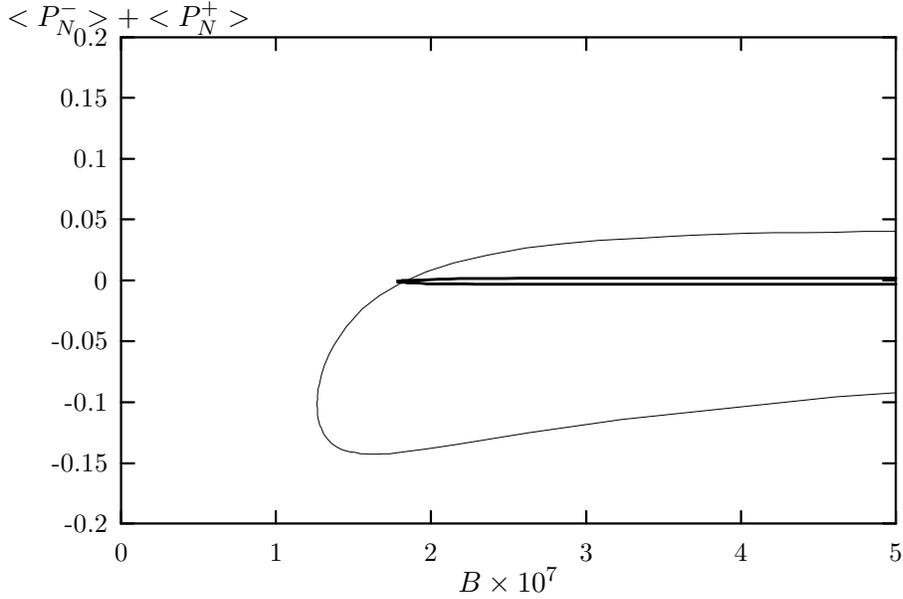


\section{Summary and Conclusions }

The most general model--independent analysis of lepton
polarization asymmetries in the rare $B$ decay $\Bstt $
is presented. We have
presented
the longitudinal, normal and transverse
polarization asymmetries of $\tau^+ $ and $\tau^- $ as
functions of the coefficients of the twelve four--Fermi operators,
ten local and two nonlocal ones.
Even though the experimental observations of all those asymmetries may
be very challenging, we found that 
such observations
will be very useful
to pin down new physics beyond the Standard Model (SM) systematically,
telling us which type of operators contributes to the process.

We also investigated 
various
combinations of the polarization asymmetries:
For the longitudinal polarization $P_L$, 
the contribution from the SM to $P_L^- $ of $\tau^-$ is exactly the same as
that to $P_L^+ $ of $\tau^+$, with just the opposite sign.  
Therefore, if there is any difference in absolute values
between the longitudinal asymmetries of $\tau^+ $ and $\tau^-$, 
it must come from interactions beyond the SM -- in this case 
from the terms with the scalar-- and tensor--type interactions,
because the contributions from the combination of 
vector types and nonlocal types are
all canceled. 
We also found that the contribution from $C_{LRLR } + C_{LRRL}$ is 
much larger than from the other scalar--type interactions. 
We also showed the usefulness of the transverse asymmetry $P_T $. 
From the difference between $P_T^- $ and $P_T^+ $, we can find the dependence 
on $C_7*C_{10} $,
if no contributions beyond those of the SM exist.
However, if there exist new physics interactions,
the contribution from the scalar--type operators,
$C_{LRLR } - C_{LRRL}$, will dominate the difference $P_T^- - P_T^+$.
Therefore, if there are the scalar--type interactions from new physics, 
we can find the interaction strength by using the results of both
$P_L^- + P_L^+ $ and  $P_T^- - P_T^+ $.
As is well known, many new physics models (for example, multi Higgs 
doublet models \cite{HZ}, SUSY \cite{KKL,CGG}, $R$-parity violation
model \cite{GN} {\it etc.}) include such scalar--type
interactions.
Concerning $< P_N^\mp >$, experimental observation may be much
more difficult than 
in the case of other asymmetries
because of small numerical value.
However, if we can measure it, 
we will be able to get very useful information
on the imaginary part of $C_T - 2 C_{TE}$. 
Of course, within the SM, $< P_N^- > + < P_N^+ > = 0$.

To summarize, we presented
the most general model-independent analysis of the lepton polarization 
asymmetries in the rare $B$ decay, $\Bstt$. 
The longitudinal, normal and transverse 
polarization asymmetries for the $\tau^+ $ and $\tau^-$, and the combinations
of them as the functions of the Wilson coefficients of  
the twelve independent four--Fermi interactions are also 
presented.
It will be  very useful to pin down new physics systematically,
once we have the experimental data with high statistics
and the deviation from the Standard Model is found. \\    

\bigskip

\centerline{\bf ACKNOWLEDGMENTS}
\medskip

We would like to thank G. Cvetic and T. Morozumi 
for careful reading of the manuscript, their
valuable comments and for suggestions. 
The work of C.S.K. was supported 
in part by KRF Non-Directed-Research-Fund, Project No. 1997-001-D00111,
in part by the BSRI Program, Ministry of Education, Project No. 98-015-D00061,
in part by the KOSEF-DFG large collaboration project, 
Project No. 96-0702-01-01-2. The work of T.Y. was supported in part by 
Grant-in-Aid for Scientific Research from the Ministry of Education,
Science and Culture of Japan and in part by JSPS Research Fellowships
for Young Scientists. 

\end{document}